\documentclass{KapProc} 

\normallatexbib

\usepackage{psfig,epsfig}
\newcommand{\lesa}{{\mbox{LES-$\alpha~$}}}
\newcommand{\be}{\begin{equation}}
\newcommand{\en}{\end{equation}}
\newcommand{\bea}{\begin{eqnarray}}
\newcommand{\ena}{\end{eqnarray}}
\newcommand{\pr}{\partial}

\newcommand{\recitwelve}{\frac{1}{12}}
\newcommand{\quarter}{\frac14}
\newcommand{\half}{\frac12}

\newcommand{\ovl}{\overline}

\newcommand{\ub}{{\overline{u}}}
\newcommand{\vb}{{\overline{v}}}
\newcommand{\pb}{{\overline{p}}}

\newcommand{\LL}{{\cal L}}
\newcommand{\HH}{{\cal H}}

\begin{document}



\articletitle[Alpha-modeling strategy for LES of turbulent mixing]
{Alpha-modeling strategy for LES of turbulent mixing}

\chaptitlerunninghead{{\mbox{LES-$\alpha$}} modeling of turbulent mixing}

\author{
Bernard J. Geurts \\ {\it{Faculty of Mathematical Sciences, 
University of Twente,
P.O. Box 217, 7500 AE Enschede, The Netherlands}} \\ 
{\sf{b.j.geurts@math.utwente.nl}}
\\ ~ \\
Darryl D. Holm \\ {\it{Theoretical Division and Center for Nonlinear Studies,
Los Alamos National Laboratory, MS B284 Los Alamos, NM 87545, USA}} 
\\ {\sf{dholm@lanl.gov}}
}



  \begin{abstract}
The $\alpha$-modeling strategy is followed to derive a new
subgrid parameterization of the turbulent stress tensor in large-eddy 
simulation (LES).
The \lesa modeling yields an explicitly filtered subgrid 
parameterization which contains
the filtered nonlinear gradient model as well as a model which represents
`Leray-regularization'. The \lesa model is compared with similarity and
eddy-viscosity models that also use the dynamic procedure. Numerical 
simulations
of a turbulent mixing layer are performed using both a second order, 
and a fourth order
accurate finite volume discretization. The Leray model emerges as the 
most accurate,
robust and computationally efficient among the three \lesa subgrid 
parameterizations for
the turbulent mixing layer. The evolution of the resolved kinetic 
energy is analyzed and
the various subgrid-model contributions to it are identified. By 
comparing \lesa at
different subgrid resolutions, an impression of finite volume 
discretization error
dynamics is obtained.
\end{abstract}



\section{Introduction}

Accurate modeling and simulation of turbulent flow is a topic of 
intense ongoing
research. The approaches to this problem area can be distinguished, 
e.g., by the amount
of detail that is intended to be included in the physical and 
numerical description.
Simulation strategies that aim to calculate the full, unsteady solution of
the governing Navier-Stokes equations are known as direct numerical 
simulations (DNS).
The DNS approach does not involve any modeling or approximation 
except its numerical
nature and in principle it can provide solutions that possess all 
dynamically relevant
flow features \cite{spalart,rogallo}. In turbulent flow, these 
features range from large,
geometry dependent scales to very much smaller dissipative 
length-scales. While accurate
in principle, the DNS approach is severely restricted by limitations 
in spatial and
temporal resolution, even with modern computational capabilities, 
because of the tendency
of fluid flow to cascade its energy to smaller and smaller scales.

This situation summons alternative, restricted simulation approaches 
to the turbulent
flow problem that are aimed at capturing the primary features of the 
flow above a certain
length-scale only. A prominent example of this is the large-eddy 
simulation (LES) strategy
\cite{lesieur}. Rather than aiming for a precise and complete 
numerical treatment of all
features that play a role in the evolution of the flow, an element of 
turbulence
modeling is involved in LES \cite{meneveau}. In the filtering 
approach to LES, this
modeling element is introduced by applying a spatially localized filter
operation to the Navier-Stokes equations \cite{germano_fil}. This introduces a
smoothing of the flow features and a corresponding reduction in the 
flow complexity
\cite{geurtscam}. One commonly adopts spatial convolution filters 
which effectively remove
the small-scale flow features that fall below an externally 
introduced length-scale
$\Delta$, referred to as the filter-width. This smoothing can 
significantly reduce the
requirements on the resolution and, thus, allow LES to be performed 
for much more
realistic situations than DNS, e.g., at higher Reynolds number, within the same
computational capabilities \cite{vreman_jfm}. This constitutes the 
main virtue of LES.

The LES approach is conceptually different from the
Reynolds Averaged Navier-Stokes (or, RANS) approach, which is based on
statistical arguments and exact ensemble averages that raise the
classic turbulence closure problem.
When the spatially localized smoothing operation in LES is applied to 
the nonlinear
convective terms in the Navier-Stokes equations, this also gives rise to
a closure problem that needs to be resolved. Thus, the LES approach 
must face its own
turbulence closure problem: How to model the effects of the 
filtered-out scales in terms
of the remaining resolved fields?

In the absence of a comprehensive theory of turbulence, empirical 
knowledge about
subgrid-scale modeling is essential but still incomplete. Since in 
LES only the dynamical
effects of the smaller scales need to be represented, the modeling is 
supposed to be
simpler and more straightforward, compared to the setting encountered in
statistical modeling such as in RANS. To guide the construction of 
suitable models we
advocate the use of constraints based on rigorous properties of the 
LES modeling problem
such as realizability conditions \cite{vreman_rea} and algebraic identities
\cite{germano_fil,geurtscam}. A thoughtful overview of these 
constraints is given in
\cite{ghosal}.

In this paper, we follow the $\alpha$-modeling approach to the LES 
closure problem.
The $\alpha$-modeling approach is based on the Lagrangian-averaged
Navier-Stokes$-\alpha$ equations (LANS$-\alpha$, or NS$-\alpha$) 
described below. The
LANS$-\alpha$ approach eliminates some of the heuristic elements that would
otherwise be involved in the modeling. The original LANS$-\alpha$ theory also
involves an elliptic operator inversion in defining its stress tensor.  When we
apply filtering in defining the LANS$-\alpha$ stress tensor, instead 
of the operator
inversion in the original theory, we call it \lesa.

\paragraph{Background and references for LANS$-\alpha$, or 
NS$-\alpha$ equations}
The inviscid LANS$-\alpha$ equations (called Euler$-\alpha$, in the 
absence of viscosity)
were introduced through a variational formulation in \cite{HMR-AIM[1998]},
\cite{HMR-PRL[1998]} as a generalization to $3D$ of the integrable inviscid
$1D$ Camassa--Holm equation discovered in \cite{CH-PRL[1993]}.  A
connection between turbulence and the solutions of the viscous $3D$ 
Camassa--Holm, or
Navier--Stokes--alpha (NS$-\alpha$) equations was identified, when 
viscosity was
introduced in \cite{Chen-etal[1998]}--\cite{Chen-etal[1999b]}. 
Specifically, the steady
analytical solution of the NS$-\alpha$ equations was found to compare 
successfully  with
experimental and numerical data for mean velocity and Reynolds 
stresses for turbulent
flows in pipes and channels over a wide range of Reynolds numbers. 
These comparisons
suggested the NS$-\alpha$ equations could be used as a closure model 
for the mean effects
of subgrid excitations. Numerical tests further substantiating this 
intuition were
performed and reported in \cite{Chen-etal[1999c]}.

An alternative more ``physical'' derivation  for the inviscid
NS$-\alpha$  equations (Euler$-\alpha$), was introduced in
\cite{Holm-PhysD[1999]} (see also \cite{Chen-etal[1999a]}). This alternative
derivation was based on substituting in Hamilton's principle the
decomposition of the Lagrangian fluid-parcel trajectory into its
mean and fluctuating components at linear order in the fluctuation 
amplitude. This was
followed by making the Taylor hypothesis for frozen-in turbulence and 
averaging at
constant Lagrangian coordinate, before taking variations. Hence, the 
descriptive name
Lagrangian-averaged Navier-Stokes$-\alpha$ equations (LANS$-\alpha$) 
was given for the
viscous version of this model. A variant of this approach was also 
elaborated in
\cite{MS-ARMA[2001]} but this resulted in a second-grade fluid model, 
instead of the
viscous LANS$-\alpha$ equations, because the choice of dissipation made in
\cite{MS-ARMA[2001]} differed from the Navier-Stokes dissipation chosen in
\cite{Chen-etal[1998]}--\cite{Holm-PhysD[1999]}.  The geometry and 
analysis of the
inviscid Euler$-\alpha$ equations was presented in \cite{Shkoller[1998]},
\cite{MRS-GFA[2000]}. The analysis of global existence and 
well-posedness for the viscous
LANS$-\alpha$ was given for periodic domains in \cite{FHT-JDE[2002]} 
and was modified for
bounded domains in \cite{MS-PhilTransRoySoc[2001]}. For more 
information and a guide to the
previous literature specifically about the NS$-\alpha$ model, see paper
\cite{FHT-PhysD[2001]}. The latter paper also discusses  connections 
to standard
concepts and scaling laws in turbulence modeling, including 
relationships of the
NS$-\alpha$ model to large eddy simulation (LES) models that are 
pursued farther in the
present paper. Related results interpreting the NS$-\alpha$ model as 
an extension of scale
similarity LES models of turbulence are also reported in 
\cite{doma_holm}. A numerical
comparison of LANS$-\alpha$ model results with LES models for the 
late stages of decaying
homogeneous turbulence is discussed in \cite{Mohseni-etal[2000]}. Vortex
interactions in the early stages of $3D$ turbulence decay are studied 
numerically with
LANS$-\alpha$ and compared with both DNS and the Smagorinsky eddy 
viscosity approach in
\cite{HK[2001]}.

\paragraph{Three contributions in the present approach}
Stated most simply, the LANS$-\alpha$ approach may be interpreted as a closure
model for the turbulent stress tensor that is derived from Kelvin's 
circulation theorem,
using a smoothed transport velocity, as discussed in \cite{FHT-PhysD[2001]},
\cite{doma_holm}, \cite{Holm-NC[1999]}. A new development within this 
approach is
introduced here that gives rise to an explicitly filtered similarity-type model
\cite{bardina} for the turbulent stress tensor, composed of three 
different contributions.
The first contribution is a filtered version of the nonlinear 
gradient model. The
unfiltered version of this model is also known as the `Clark' model 
\cite{leonard,clark},
the `gradient' model \cite{vreman_tcfd} or the `tensor-diffusivity' model
\cite{winckelmans}. The second contribution, when combined with the
filtered nonlinear gradient model, represents the so-called `Leray 
regularization' of
Navier-Stokes dynamics \cite{leray}. Finally, a new third 
contribution emerges from
the derivation which completes the full \lesa model and endows it 
with its own Kelvin's
circulation theorem.

To investigate the physical and numerical properties of the resulting
three-part LANS$-\alpha$ subgrid parameterization, we consider a turbulent
mixing layer \cite{vreman_jem}. This flow is well documented in 
literature and provides a
realistic canonical flow problem \cite{debruin} suitable for testing 
and comparison with
predictions arising from more traditional subgrid model developments 
\cite{vreman_jfm}. In
particular, we consider similarity, and eddy-viscosity modeling, 
combined with the
dynamic procedure based on Germano's identity 
\cite{germano_fil,geurtscam,germano_iden},
to compare with \lesa. In addition to the full \lesa model, in our 
comparisons we also
consider the two models that are contained in it, i.e., the filtered 
nonlinear gradient
model and the Leray model. We will refer to all three as \lesa models.
For all these models, the explicit filtering stage is essential. Without this
filtering operation in the definition of the models, a finite time 
instability is observed
to arise in the simulations. The basis for this instability can be 
traced back to the
presence of antidiffusion in the nonlinear gradient contribution. We 
sketch an analysis
of the one-dimensional Burgers equation, following 
\cite{vreman_tcfd}, to illustrate this
instability and show, through simulation, that increasing the subgrid 
resolution further
enhances this instability. Analyzing the resolved kinetic energy 
dynamics reveals that
this instability  is associated with an excessive contribution to back-scatter.

The `nonlinearly dispersive' filtered models that arise in \lesa  are 
reminiscent of
similarity LES models \cite{doma_holm}. The \lesa model separates the
resolved kinetic energy (RKE) of the flow into the sum of two 
contributions: namely, the
energies due to motions at scales that are either greater, or less 
than an externally
determined length-scale ($\alpha$). The two contributions are modeled by
\begin{equation}
RKE = RKE^{(>)} + RKE^{(<)}
\end{equation}
As we shall describe later in reviewing the \lesa strategy, the 
kinetic energy $RKE^{(<)}$
of turbulent motions at scales less than $\alpha$ is modeled by a 
term proportional to the
{\it rate of dissipation} of the kinetic energy $RKE^{(>)}$ at scales 
greater than
$\alpha$. (The time-scale in the proportionality constant is the 
viscous diffusion time
$\alpha^2/2\nu$.)

A key aspect of the \lesa dynamics is the exchange, or conversion, of 
kinetic energy
between $RKE^{(>)}$ and $RKE^{(<)}$. We focus on the contributions to 
the dynamics of the
resolved kinetic energy $RKE^{(>)}$ at scales greater than $\alpha$ 
that arise from the
different terms in the \lesa models. The filtered nonlinear gradient 
model, the Leray model
and the full \lesa model all contribute to the reduction of the 
$RKE^{(>)}$ in the laminar
stages of the flow. This corresponds to forward scatter of $RKE^{(>)}$ into 
$RKE^{(<)}$ outweighing backward scatter. In the
developing turbulent flow regime, the  resolved kinetic energy 
$RKE^{(>)}$ of the full
\lesa model may decrease too slowly compared to DNS, and for some 
settings of the
(numerical) parameters can even become reactive in nature, thereby 
back-scattering too much kinetic
energy from $RKE^{(<)}$ into $RKE^{(>)}$. In contrast, the 
contribution of the Leray model
to the $RKE^{(>)}$ dynamics remains forward in nature and appears to 
settle around some
negative, nonzero value in the turbulent regime. All the \lesa models 
show contributions
to both forward and backward scatter of RKE. It is observed that in 
the full \lesa
model two of the three terms almost cancel in the evolution of resolved
kinetic energy $RKE^{(>)}$. This cancellation nearly reduces the full \lesa
model to the filtered nonlinear gradient model.

The mixing layer simulations indicate that the Leray subgrid model 
provides more accurate
predictions compared to both the filtered nonlinear gradient and the 
full \lesa model.
This is based on comparisons that include mean flow quantities, 
fluctuating flow
properties and the energy spectrum. In addition, the Leray \lesa 
model appears more robust
with respect to changes in numerical parameters. Predictions based on 
this model compare
quite favorably with those obtained using dynamic (mixed) models and 
filtered DNS
results.  The Leray model combines this feature with a strongly 
reduced computational cost
and is favored for this reason, as well. In addition, a number of classic
mathematical properties (e.g., existence and uniquenes of strong 
solutions) can be proven
rigorously for fluid flows that are modeled with Leray's 
regularization. These can be used
to guide further developments of this model such as extensions to 
more complex flows at
higher Reynolds number. This is a topic of current research and will 
be published elsewhere
\cite{geurts_holm}.

Apart from the problem of modeling the subgrid-scale stresses, any 
actual realization of
LES is inherently endowed with (strongly) interacting errors arising 
from the required
use of marginal numerical resolution
\cite{ghosal_numer,vreman_numer1,vreman_numer2,geurts_froehlich}.
The accuracy of the predictions depends on the numerical method and 
subgrid resolution
one uses. We consider in some detail numerical contamination of a 
`nonlinear gradient
fluid' and a `Leray fluid,' which are defined as the hypothetical 
fluids governed by the
corresponding subgrid model. In this analysis we are consequently not 
concerned with how
accurately the modeled equations represent filtered DNS results. 
Rather, we focus
on the numerical contamination of the predictions. For this purpose 
we compare two finite
volume spatial discretization methods, one at second order, and the 
other at fourth order
accuracy.

The subgrid modeling and the spatial discretization of the equations 
give rise to a
computational dynamical system whose properties are intended to 
simulate those of the
filtered Navier-Stokes equations. The success of this simulation 
depends of course on the
properties of the model, as well as of the spatial discretization 
method and the subgrid
resolution. The model properties are particularly important in view 
of the marginal
subgrid resolution used in present-day LES. We consider the role of 
the numerical method
at various resolutions and various ratios of the filter-width 
$\Delta$ compared to the
grid-spacing $h$. Let $\Delta$ be a fixed constant. In cases of large  ratios
$\Delta/h\gg1$ one approximates the grid-independent LES solution 
corresponding to the
given value of $\Delta$, and the accuracy of its predictions will be 
limited by the quality
of the assumed subgrid model. At the other extreme, one may assume 
$\Delta/h$ to be rather
small and numerical effects can constitute a large source of error. 
Through a systematic
variation of the ratio $\Delta/ h$ at constant $\Delta$ we can 
identify the contributions
of the numerical method at coarse resolutions. This will give an 
impression of how the
computational dynamical system is affected by variations in the 
resolution and the
numerical method.

The organization of this chapter is as follows. In section~\ref{les} 
we introduce the
large-eddy simulation problem and identify the closure problem and 
some of its properties.
The treatment of this closure problem using the $\alpha$-framework is 
sketched, together
with more conventional subgrid parameterization that involves the 
introduction of
similarity, and eddy-viscosity modeling. Finally, we analyze the instabilities
associated with the use of the unfiltered nonlinear gradient model. 
In section~\ref{sim}
we introduce the numerical methods used and consider the simulation 
of a turbulent mixing
layer. Some direct and large-eddy simulation results will be shown. In
section~\ref{lesmxl} we focus on the \lesa models and consider the 
dynamics of the
resolved kinetic energy in each of the three cases. This comparison 
provides a framework
for understanding how the different \lesa subgrid models function. We 
proceed with an
assessment of the numerical error dynamics at relatively coarse 
subgrid resolutions.
A summary and concluding remarks for the chapter are given in 
section~\ref{concl}.

\section{Large-eddy simulation and $\alpha$-modeling}
\label{les}

This section sketches the traditional approach to large-eddy simulation,
which arises from direct spatial filtering of the Navier-Stokes equations
(section~\ref{sec2-1}). The algebraic and analytic properties of the 
LES modeling
problem will be discussed first. The LES closure problem will then be 
considered in the
$\alpha$-framework of turbulent flow, derived via Kelvin's circulation theorem
for a smoothed, spatially filtered transport velocity 
(section~\ref{sec2-2}). The
closure of the filtered fluid flow problem achieved this way will be 
compared with the
more traditional methods of similarity, and eddy-viscosity modeling 
for LES. The latter
is introduced in section~\ref{sec2-3} together with the dynamic 
procedure based on
Germano's identity. We also sketch a stability analysis of the one-dimensional
filtered Burgers equation involving the nonlinear gradient subgrid 
model that illustrates
the instabilities associated with this model (section~\ref{sec2-4}).

\subsection{Spatially filtered fluid dynamics}
\label{sec2-1}

We consider the incompressible flow problem in $d$ spatial 
dimensions. The Cartesian
velocity fields $u_i$ ($i=1, \ldots, d$) and the normalized pressure 
field $p$ constitute
the complete solution. The velocity field is considered to be 
solenoidal and the evolution
of the solution is described by the Navier-Stokes equations. These 
are conservation laws
for mass and momentum, respectively, that can be written in the 
absence of forcing as
\bea
\pr_{j}u_{j}&=&0 \label{incom} \\
\pr_{t}u_{i}+\pr_{j}(u_{i}u_{j})+\pr_{i}p -\frac{1}{Re}\pr_{jj}u_{i}&=&0
\label{nseqs}
\ena
where $\pr_{t}$ and $\pr_{j}$ denote, respectively, partial 
differential operators in time $t$
and Cartesian coordinate $x_{j}$, $j=1, \ldots, d$. The quantity $Re=u_r \l_r/
\nu_r$ is the Reynolds number based on reference velocity ($u_r$), 
reference length
($\l_r$) and reference kinematic viscosity ($\nu_r$), which were selected to
non-dimensionalize the governing equations. Repeated indices are 
summed over their range,
except where otherwise noted.

Equations~(\ref{incom} - \ref{nseqs}) model incompressible flow in 
all its spatial
and temporal details. In deriving approximate equations that are 
specialized to capture
the generic large-scale flow features only, one applies a spatial 
filter operation
$L: u \rightarrow \ub$ to~(\ref{incom} - \ref{nseqs}).
For simplicity, we restrict to linear convolution filters:
\be
\ub({\bf x},t)=L(u)({\bf x},t)=\int_{-\infty}^{\infty} G({\bf 
x}-\mbox{\boldmath $\xi$})~
u(\mbox{\boldmath $\xi$},t)~ d \mbox{\boldmath $\xi$} = \Big( G * u 
\Big)({\bf x},t)
\en
in which the filter-kernel $G$ is normalized, i.e., $L(c)=c$ for any 
constant solution
$u=c$. We assume that the filter-kernel $G$ is localized as a function of ${\bf
x}-\mbox{\boldmath $\xi$}$ and a filter-width $\Delta$ can be 
assigned to it. Typical
filters which are commonly considered in LES are the top-hat, the 
Gaussian and the
spectral cut-off filter. Here, we restrict ourselves to the top-hat 
filter which has a
filter-kernel given by
\be
G({\bf z})=\left\{\begin{array}{ll}
               \Delta^{-3} & \mbox{if
                $|z_{i}|<\Delta_i/2$} \\
                0 & \mbox{otherwise}
            \end{array}
            \right.
\en
where $\Delta_{i}$ denotes the filter-width in the $x_{i}$ direction 
and the total
filter-width $\Delta$ is specified by
\be
\Delta^{3}=\Delta_{1}\Delta_{2}\Delta_{3}
\en
Apart from the filter-kernel in physical space, the Fourier-transform 
of $G({\bf z})$,
denoted by $H({\bf k})$, is important, e.g., for the interpretation 
of the effect of the
filter-operation on signals which are composed of various length-scales. The
Fourier-transform of the top-hat filter is given by (no sum in $\Delta_i k_i$)
\be
H({\bf k})=\prod_{i=1}^3 \frac{\sin(\Delta_i k_i/2)}{\Delta_i k_i/2}
\en
If we consider a general Fourier-representation of a solution $u({\bf x},t)$,
\be
u({\bf x},t)=\sum_{{\bf k}} c_{{\bf k}}(t) e^{ i {\bf k} \cdot {\bf x}}
\en
the filtered solution can directly be written as
\be
\ub({\bf x},t) =\sum_{{\bf k}} \Big( H({\bf k})c_{{\bf k}}(t) \Big) 
e^{ i {\bf k} \cdot
{\bf x}}
\en
We notice that each Fourier-coefficient $c_{{\bf k}}(t)$ is 
attenuated by a factor
$H({\bf k})$. The normalization condition of the filter-operation 
implies \mbox{$H({\bf
0})=1$}. For small values of $|\Delta_{i}k_{i}|$ one infers from a 
Taylor expansion
\begin{equation}
H({\bf k})=1-(1/24)\Big((k_{1}\Delta_{1})^{2}+(k_{2}\Delta_{2})^{2}
+(k_{3}\Delta_{3})^{2} \Big) + \ldots
\label{taylor}
\end{equation}
which shows the small attenuation of flow features which are considerably
larger than the filter-width $\Delta$, i.e., $|\Delta_{i}k_{i}| \ll 
1$ for $i=1,~2,~3$.
As $|{\bf k}|$ increases $H({\bf k})$ becomes smaller while 
oscillating as a function of
$\Delta_{i}k_{i}$. Consequently, the coefficients $c_{{\bf k}}(t)$ 
are strongly reduced as
$|\Delta_{i}k_{i}| \gg 1$ and the small scale features in the 
solution are effectively
taken out by the filter operation. Similarly, the Gaussian filter can 
be shown to have the
same expansion for small $|\Delta_{i}k_{i}|$ and reduces to zero 
monotonously as
$|\Delta_{i}k_{i}|$ becomes large.

The filter operation $L$ is a convolution integral. Hence, it is a 
linear operation that
commutates with partial derivatives \cite{geurts_kreta,ghosal_moin}. 
This property
facilitates the application of the filter to the governing 
equations~(\ref{incom} -
\ref{nseqs}). A straightforward application of such filters leads to
\bea
\pr_{j}\ub_{j}&=&0 \label{incomles} \\
\pr_{t}\ub_{i}+\pr_{j}(\ub_{i}\ub_{j})+\pr_{i}\pb -\frac{1}{Re}\pr_{jj}\ub_{i}
&=& -\pr_{j}\tau_{ij}
\label{leseqs}
\ena
where we introduced the turbulent stress tensor
\be
\tau_{ij}={\overline{u_{i}u_{j}}}-\ub_{i} \ub_{j}
\en
We observe that the filtered solution $\{ \ub_{i},~\pb \}$ represents 
an incompressible
flow ($\pr_{j}\ub_{j}=0$). The same differential operator as in (\ref{nseqs})
acts on $\{ \ub_{i},~\pb \}$ and due to the filtering a non-zero 
right-hand side has
arisen which contains the divergence of the turbulent stress tensor 
$\tau_{ij}$. This
latter term is the so-called subgrid term, and expressing it in terms 
of the filtered
velocity and its derivatives constitutes the closure problem in 
large-eddy simulation.

The LES modeling problem as expressed above has a number of important, rigorous
properties which may serve as guidelines for specifying appropriate 
subgrid-models for
$\tau_{ij}$. In particular, we will briefly review realizability 
conditions, algebraic
identities and transformation properties. Adhering to these basic features of
$\tau_{ij}$ limits some of the heuristic elements in the subgrid-modeling.

\subsubsection{Realizability}

It is well known that the Reynolds stress $\ovl{u_{i}'u_{j}'}$ in 
RANS is positive
semi-definite \cite{vachat,schumann} and the following inequalities hold 
\cite{ortega}
\bea
\tau_{ii}&\geq& 0 ~~~~~~~~~~~~\mbox{for}~~~~ i\in\{1,2,3\} \quad\hbox{(no sum)}
\label{A1} \\
|\tau_{ij}|&\leq& \sqrt{\tau_{ii} \tau_{jj}}
~~~~\mbox{for}~~~~ i,j\in\{1,2,3\}
\quad\hbox{(no sum)}
\label{A2} \\
\mbox{det}(\tau_{ij}) &\geq& 0
\ena
If the filtering approach is followed, in general $\tau_{ij}\neq
\ovl{u_i' u_j'}$ and, therefore, it is relevant to know the conditions
under which $\tau_{ij}$ is positive semi-definite.
Following Vreman {\it  et al.} \cite{vreman_rea}, it can be proved 
that $\tau_{ij}$ in LES
is positive semi-definite if and only if the filter kernel
$G({\bf x},\mbox{\boldmath $\xi$})$ is positive for all ${\bf x}$ and 
$\mbox{\boldmath
$\xi$}$. If we assume $G\geq 0$, the expression
\be
(f,g)=\int_{\Omega} G({\bf x},\mbox{\boldmath $\xi$})
f(\mbox{\boldmath $\xi$})g(\mbox{\boldmath $\xi$})d\mbox{\boldmath $\xi$}
\en
defines an inner product and we can rewrite the turbulent
stress tensor as:
\be
\tau_{ij}({\bf x}) = \int_{\Omega}G({\bf x},\mbox{\boldmath $\xi$})
(u_i(\mbox{\boldmath $\xi$})-\ovl{u}_i({\bf x}))
  (u_j(\mbox{\boldmath $\xi$})-\ovl{u}_j({\bf x})) d\mbox{\boldmath $\xi$}
  = (v_i^{\bf x},v_j^{\bf x})
\label{inttau}
\en
with $v_i^{\bf x}(\mbox{\boldmath $\xi$}) \equiv u_i(\mbox{\boldmath
$\xi$})-\ovl{u}_i({\bf x})$. In this way the tensor $\tau_{ij}$ forms 
a $3 \times 3$
Grammian matrix of inner products. Such a matrix is always positive 
semi-definite and
consequently $\tau_{ij}$ satisfies the realizability conditions.  The 
reverse statement
can likewise be established, showing that the condition $G \geq 0$ is 
both necessary and
sufficient.

One prefers the turbulent stress tensor $\tau_{ij}$ in LES
to be realizable for a number of reasons. For example, if
$\tau_{ij}$ is realizable, the generalized turbulent kinetic energy
$k=\tau_{ii}/2$ is a positive quantity. This quantity is required to 
be positive
in subgrid models which involve the $k$-equation 
\cite{ghosal_lund_moin_akselvoll}.
Several further benefits of realizability and positive filters can be 
identified
\cite{vreman_rea}; here we restrict to adding that the kinetic energy 
of ${\overline{u}}$
is bounded by that of $u$ for positive filter-kernels:
\be
\frac12 \int_{\Omega} |{\overline{u}}|^{2} d{\bf x} \leq
\frac12 \int_{\Omega} |u|^{2} d{\bf x}
\en

Requiring realizability places some restrictions on subgrid models. For
example, if $G \geq 0$  models for $\tau$ should be
realizable. Consider, e.g., an eddy-viscosity model
$m_{ij}$ given by
\be
m_{ij}=-\nu_{e} \sigma_{ij}+\frac{2}{3}k\delta_{ij}
\en
In order for this model to be realizable, a lower bound
for $k$ in terms of the eddy-viscosity $\nu_{e}$ arises, i.e.,
$k \geq \frac{1}{2}\sqrt{3\sigma} \nu_{e}$
where $\sigma=\frac12 \sigma_{ij}\sigma_{ij}$ and $\sigma_{ij}$ is the
rate of strain
tensor given by $\sigma_{ij}=\pr_{i}u_{j}+\pr_{j}u_{i}$.

\subsubsection{Algebraic identities}

The introduction of the
product operator $S(u_{i},u_{j})=u_{i}u_{j}$ allows to write the
turbulent stress tensor as \cite{geurtscam}:
\be
\tau_{ij}^{L}={\overline{u_{i}u_{j}}}
-{\overline{u}}_{i}{\overline{u}}_{j}
= L(S(u_{i},u_{j}))-S(L(u_{i}),L(u_{j}))=
[L,S](u_{i},u_{j})
\en
in terms of the central commutator $[L,S]$ of the filter $L$
and the product operator $S$. This
commutator shares a number of properties with the Poisson-bracket in classical
mechanics. Leibniz' rule of Poisson-brackets is in the context
of LES known as Germano's identity \cite{germano_fil}
\begin{equation}
\label{basicid}
[ \LL_{1} \LL_{2},S]=[\LL_{1},S]\LL_{2}+\LL_{1} [\LL_{2},S]
\end{equation}
This can also be written as
\begin{equation}
\tau^{\LL_{1} \LL_{2}} = \tau^{\LL_{1}} \LL_{2}
+ \LL_{1} \tau^{\LL_{2}}
\label{usual}
\end{equation}
and expresses the relation between the turbulent stress tensor corresponding
to different filter-levels. In these identities,
$\LL_{1}$ and $\LL_{2}$ denote any two filter operators and
$\tau^{K}=[K,S]$. The first term on
the right-hand side of (\ref{usual}) is interpreted as the `resolved' 
term which
in an LES can be evaluated without further approximation. The other two terms
require modeling of $\tau$ at the corresponding filter-levels.

Similarly, Jacobi's identity holds for $S$, $\LL_{1}$ and
$\LL_{2}$:
\begin{equation}
\label{jacobi}
[ \LL_{1},[\LL_{2},S]]+[\LL_{2},[S,\LL_{1}]]=-[S,[\LL_{1},\LL_{2}]]
\end{equation}
The expressions in (\ref{basicid}) and (\ref{jacobi}) provide relations
between the turbulent stress tensor corresponding to different filters
and these can be used to dynamically model $\tau^{L}$. The success of models
incorporating Germano's identity  (\ref{basicid}) is by now well established in
applications for many different flows. In the traditional formulation 
one selects
$\LL_{1}=\HH$ and $\LL_{2}=L$ where $\HH$ is the so called 
test-filter. In this case one
can specify Germano's identity \cite{germano_fil} as
\begin{equation}
  \tau^{\HH L}(u) = \tau^{\HH} \left( L (u) \right) +
\HH \left( \tau^{L} (u) \right)
\label{tradgerm}
\end{equation}
The first term on the right hand side involves the operator $\tau^{\HH}$ acting
on the resolved LES field $L(u)$ and during an LES this is known
explicitly. The remaining terms need to be replaced by a model. In the
dynamic modeling \cite{germano_iden}
the next step is to assume a base-model $m^{K}$ corresponding to
filter-level $K$ and optimize any coefficients in it,  e.g., in a least
squares sense \cite{lilly}. The operator formulation can be extended to include
approximate inversion defined by ${\cal L}^{-1}(L(x^{k}))=x^{k}$ for 
$0 \leq k \leq N$
\cite{geurts97}. Dynamic inverse models have been applied in mixing 
layers \cite{kuerten}.

\subsubsection{Transformation properties}

The turbulent stress tensor $\tau_{ij}$ can be shown to be invariant 
with respect
to Galilean transformations. This property also holds for the divergence,
i.e., $\pr_{j}\tau_{ij}$, referred to as the subgrid scale force. 
Hence, the filtered
Navier-Stokes dynamics is Galilean invariant. Suitable subgrid models 
should at least
maintain the Galilean invariance of the divergence of the model, 
i.e., $\pr_{j}m_{ij}$. In
fact, most subgrid models are represented by tensors which are 
Galilean invariant.
Examples of non-symmetric tensor models have been reported in 
\cite{doma_holm} for which, however, $\pr_{j}m_{ij}$ was verified 
to be Galilean
invariant.

Likewise, it is of interest to consider a transformation of the 
subgrid scale stress
tensor to a frame of reference rotating with a uniform angular 
velocity. The full subgrid
scale stress tensor transforms in such a way that the subgrid scale 
force is the same in
an  inertial and in a  rotating frame. Horiuti~\cite{horiuti00}
has recently analyzed several subgrid scale models and showed that 
some of them do
not satisfy this condition. This is an example of how transformation 
properties of
the exact turbulent stress tensor can be used to guide
propositions for subgrid modeling.

\vspace*{2mm}

After closing the filtered equations (\ref{incomles}-\ref{leseqs}) by 
a subgrid model
stress tensor $m_{ij}$ we arrive at the modeled filtered dynamics 
described in the
absence of forcing  by
\bea
\pr_{j}v_{j}&=&0\\
\pr_{t}v_{i}+\pr_{j}(v_{i}v_{j})+\pr_{i}P 
-\frac{1}{Re}\pr_{jj}v_{i}&=& -\pr_{j}m_{ij}
\label{vleseq}
\ena
whose solution is denoted as $\{v_{i},~P\}$. Ideally, if $m_{ij}$ and the
numerical treatment were correct and had no undesirable effects on 
the dynamics,
one might expect $v_{i}=\ub_{i}$. In view of possible sensitive 
dependence of an actual
solution,  e.g., on the initial condition, one should not expect 
instantaneous and
point-wise equality of $v_{i}$ and $\ub_{i}$ but rather one should 
expect statistical
properties of the filtered and modeled solution to be equal. 
Assessing the extent to
which the properties of $\{ v_{i}, P \}$ and $\{ \ub_{i}, \pb \}$ are 
correlated allows
an evaluation of the quality of the subgrid model, the dynamic 
effects arising from
the numerical method and the interactions between modeling and 
numerics. In what follows,
we will use the notation $\{ v_{i}, P \}$ to distinguish the solution 
of the subgrid model
from the filtered solution $\{ \ub_{i}, \ovl{p} \}$.

\subsection{Subgrid model derived from Kelvin's theorem}
\label{sec2-2}

The \lesa modeling scheme we shall use here is based on the well-known
viscous Camassa-Holm equations, or LANS$-\alpha$ model. This modeling 
strategy imposes a
``cost" in resolved kinetic energy (RKE) for creation of smaller and 
smaller excitations
below a certain, externally specified length scale, denoted by 
$\alpha$. This cost in
converting $RKE^{(>)}$ to $RKE^{(<)}$ implies a nonlinear  modification of the
Navier-Stokes equations which is reactive, or dispersive, in nature 
instead of being
diffusive, as is more common in present-day LES modeling. The 
modification appears in the
nonlinear convection term and can be rewritten in terms of a subgrid 
model for the
turbulent stress tensor. In the LANS$-\alpha$ model, the processes of 
nonlinear conversion
of $RKE^{(>)}$ to $RKE^{(<)}$ and sweeping of the smaller scales by 
the larger ones are
still included in the modeled dynamics.  We will sketch the \lesa 
approach in this
subsection and extract the subgrid models used in this study. For 
more details and
applications of this approach, see
\cite{Chen-etal[1998]}--\cite{Holm-PhysD[1999]},
\cite{FHT-PhysD[2001]}--\cite{HK[2001]}.

It is well known that the Navier-Stokes equations satisfy
Kelvin's circulation theorem, i.e.,
\begin{equation}\label{KelThm-NS}
\frac{d}{dt}
\oint_{\gamma(\mathbf{u})} ~u_{j} ~dx_{j} =
\oint_{\gamma(\mathbf{u})} ~\frac{1}{Re}~ \pr_{kk} u_{j} ~dx_{j}
\end{equation}
Here $\gamma(\mathbf{u})$ represents a fluid loop that moves with the 
Eulerian fluid
velocity $\mathbf{u}(\mathbf{x},t)$. The basic equations in the \lesa 
modeling may be
introduced by modifying the velocity field by which the fluid loop is 
transported. The
governing \lesa equations will provide the smoothed solution $\{ 
v_{j},~P \}$ and we
specify the equations for ${\bf v}$ through the Kelvin-filtered 
circulation theorem.
Namely, we integrate an approximately `defiltered' velocity ${\mathbf{w}}$
around a loop $\gamma({\mathbf{v}})$ that moves with the regularized 
spatially filtered
fluid velocity ${\mathbf{v}}$, cf. \cite{FHT-PhysD[2001]}, \cite{doma_holm},
\cite{Holm-NC[1999]}
\begin{equation}\label{KelThm-NS-alpha}
\frac{d}{dt}
\oint_{\gamma(\mathbf{v})} ~{w}_{j} ~dx_{j}
  =
\oint_{\gamma(\mathbf{v})} ~\frac{1}{Re}~ \pr_{kk} {w}_{j} ~dx_{j}
\end{equation}
Hence, the basic transport properties of the \lesa model arise from 
filtering the
`loop-velocity' to obtain $\mathbf{v}$, then approximately 
defiltering $\mathbf{v}$ to
obtain the velocity ${\mathbf{w}}$ in the Kelvin integrand. As we shall show,
this approach will yield the model stress tensor
$m_{ij}$ needed to complete the filtering approach outlined in 
section~\ref{sec2-1}.
Direct calculation of the time derivative in this modified 
circulation theorem yields the
Kelvin-filtered Navier-Stokes equations,
\begin{equation}\label{NS-alpha-eqns}
\pr_{t}{w}_{i} + v_{j} \pr_{j} {w}_{i}
+ {w}_{k} \pr_{i} v_{k} + \pr_{i} \hat{P} - \frac{1}{Re} \pr_{jj} {w}_{i}=0~,~~
\pr_{j} v_{j}= 0
\end{equation}
where we introduce the scalar function $\hat{P}$ in removing the loop integral.
The relation between the `defiltered' velocity components ${w}_i$ and 
the \lesa velocity
components $v_{i}$ of the Kelvin loop needs to be specified 
separately. The Helmholtz
defiltering operation was introduced in \cite{HMR-AIM[1998]}, 
\cite{HMR-PRL[1998]} for
this purpose:
\begin{equation}
{w}_i= v_i - \alpha^2 \pr_{jj} v_i=(1-\alpha^{2} \pr_{jj}) v_{i}
= {\cal H}_{\alpha} (v_{i})
\label{eq:hatv}
\end{equation}
where  ${\cal H}_{\alpha}$ denotes the Helmholtz operator.
We recall that all explicit filter operations $L$ with a non-zero 
second moment, have a
Taylor expansion whose leading order terms are of the same form as 
(\ref{eq:hatv}).
Consequently, we infer that the leading order relation between $\alpha$ and
$\Delta$ follows as $\alpha^{2}=\Delta^{2}/24$ for the top-hat and 
the Gaussian filter.
We will use this as the definition of $\alpha$ in the sequel.

The \lesa equations can be rearranged into a form similar to the 
basic LES equations
(\ref{vleseq}), by splitting off a subgrid model for the turbulent 
stress tensor. For the
Helmholtz defiltering, we obtain from (\ref{NS-alpha-eqns}):
\be
\pr_{t} v_i + \pr_{j} (v_i v_j) + \pr_{i} P +\pr_{j} m_{ij}^{\alpha}
- \frac{1}{Re} \pr_{jj} v_{i}
=0 ~,~~ \pr_{j}v_{j}=0
\label{eq:alpha3}
\en
after absorbing gradient terms into the redefined pressure $P$. Thus, 
we arrive at the
following parameterization for the turbulent stress tensor
\be
{\cal H}_{\alpha}(m_{ij}^{\alpha})
=
\alpha^{2}\Big(\pr_{k}v_{i}~\pr_{k}v_{j}+\pr_{k}v_{i}~ \pr_{j}v_{k}
-\pr_{i}v_{k}~\pr_{j}v_{k} \Big)
\label{alpha3-model}
\en
In the evaluation of the \lesa dynamics in the above formulation, an 
inversion of the
Helmholtz operator ${\cal H}_{\alpha}$ is required. The `exponential' (or `Yukawa') filter
\cite{germano_exp_fil} is the exact, explicit filter which inverts 
${\cal H}_{\alpha}$.
Thus, an inversion of ${\cal H}_{\alpha}$ corresponds to applying the 
exponential filter
to the right-hand side of (\ref{alpha3-model}) in order to find 
$m_{ij}^{\alpha}$.
However, since the Taylor expansion of the exponential filter is 
identical at quadratic
order to that of the top-hat and the Gaussian filters, we will 
approximate the inverse of
${\cal H}_{\alpha}$ by an application of the explicit top-hat filter, 
for reasons of
computational efficiency. Moreover, in actual simulations the 
numerical realization of the
exponential filter is only approximate and can just as well be 
replaced by the numerical
top-hat filter. This issue of (approximately) inverting the Helmholtz operator will
be studied separately and published elsewhere \cite{geurts_holm}.

The full \lesa subgrid model $m_{ij}^{\alpha}$ has three distinct 
contributions. The first
term on the right-hand side is readily recognized as the nonlinear 
gradient model which we
will denote by $A_{ij}$. This term is closely related to the 
similarity model proposed by
Bardina \cite{bardina}, as will be shown in the next subsection. The 
second term will be
denoted by $B_{ij}$ and combined with the first term, corresponds to the Leray
regularization of the convective terms in the Navier-Stokes 
equations. This regularization
arises if the familiar contribution $u_{j}\pr_{j}u_{i}$ in the 
Navier-Stokes equations is
replaced by $\vb_{j}\pr_{j}v_{i}$ in the smoothed description. The 
third term will be
denoted by $C_{ij}$. Further details of the derivation and 
mathematical properties of the
\lesa model will be published elsewhere \cite{geurts_holm}. We can 
explicitly write the
stress tensor for the \lesa model as
\begin{eqnarray} \label{alpha3_expl}
m_{ij}^{\alpha}
&=&
\frac{\Delta^{2}}{24} \Big( \ovl{\pr_{k}v_{i}~\pr_{k}v_{j}}+\ovl{\pr_{k}v_{i}~
\pr_{j}v_{k}} -\ovl{\pr_{i}v_{k}~\pr_{j}v_{k}} \Big) \nonumber \\
&=& \ovl{A_{ij}}+\ovl{B_{ij}}-\ovl{C_{ij}}
\end{eqnarray}
The explicit filter, represented by the overbar in this expression, 
is realized by the
numerical top-hat filter in this study. It does not necessarily have 
to coincide with the
LES-filter. While the LES-filter specifies the relation between the 
Navier-Stokes solution
$u_{i}$ and the \lesa solution $v_{i}$, the explicit \lesa filter is 
used to approximate
${\cal H}_{\alpha}^{-1}$. We will consider the effects associated 
with variations in
the filter-width ${\tilde{\Delta}}$ of the explicit \lesa filter with 
filter-width
${\tilde{\Delta}} / \Delta=\kappa$. Typical values that will be 
considered are $\kappa=1$
and $\kappa=2$.

In the next subsection we will describe some familiar subgrid models 
used in LES which
are based on the similarity and eddy-viscosity concepts.

\subsection{Similarity modeling and eddy-viscosity regularization}
\label{sec2-3}

We distinguish two main contributions in present-day traditional 
subgrid modeling of the
turbulent stress tensor, i.e., dissipative and similarity subgrid 
models. In this
subsection we briefly describe these two basic approaches, as well as 
subgrid models that
consist of combinations of an eddy-viscosity and a similarity part, 
so-called mixed
models. The relative importance of the two components in such mixed 
models is obtained by
using the dynamic procedure which is based upon Germano's identity 
(\ref{tradgerm}). This
mixed approach effectively regularizes and stabilizes similarity models.

As a result of the filtering, flow features of length-scales (much) 
smaller than the
filter-width $\Delta$ are considerably attenuated. This implies that 
the natural molecular
dissipation arising from the viscous fluxes, is strongly reduced, 
compared to the
unfiltered flow-problem. In order to compensate for this, dissipative 
subgrid-models have
been introduced to model the turbulent stress tensor. The prime example of such
eddy-viscosity models is the Smagorinsky model \cite{rogallo,smagorinsky}:
\be
m^{S}_{ij}=- (C_S \Delta)^2 |\sigma({\bf v})|
\sigma_{ij}({\bf v}) ~~~~\mbox{with}~~~~ |\sigma({\bf v})|^2
=\half \sigma_{ij}({\bf v}) \sigma_{ij}({\bf v})
\label{smag}
\en
where $\sigma_{ij}$ is the strain rate, introduced above
($\sigma_{ij}=\pr_{i}v_{j}+\pr_{j}v_{i}$).
This model adds only little computational overhead. The
major short-coming of the Smagorinsky model is its excessive
dissipation in laminar regions with mean shear, because $\sigma_{ij}$ is
large in such regions \cite{germano_iden}.
Furthermore, the correlation between the Smagorinsky model and
the actual turbulent stress is quite low (reported to be $\approx0.3$ 
in several flows).

In trying to compensate for these short-comings of the Smagorinsky model,
a second main branch of subgrid models emerges from the similarity 
concept \cite{bardina}.
Using the commutator notation, the turbulent stress tensor can be 
expressed as $\tau_{ij}
({\bf{u}})= [L,S](u_{i},u_{j})$. In terms of this short-hand 
notation, the basic
similarity model can be written as
\be
m^{B}_{ij}=[L,S](v_{i},v_{j})
\label{bardin}
\en
i.e., directly following the definition of the turbulent stress 
tensor, but expressed in
terms of the available modeled LES velocity field. Generalizations of 
this similarity
model arise by replacing $v_{i}$ in (\ref{bardin}) by an 
approximately defiltered field
${\widehat{v}}_{i}={\cal L}({v_{i}})$ where ${\cal L}(L(u)) \approx 
u$, i.e., ${\cal L}$
approximates the `inverse' of the filter $L$ \cite{geurts97}. In 
detail, a generalized
similarity model arises from $m^{G}= [L,S]\left( {\cal L}^{-1}(v) 
\right)$ using the
approximate inversion.  This approach is also known as the deconvolution model
\cite{stolz_adams} and is reminiscent to the subgrid estimation model
\cite{domaradzki_etal}. The correlation with $\tau_{ij}$ is much 
better with correlation
coefficients reported in the range 0.6 to 0.9 in several flows. The 
low level of
dissipation associated with these models renders them quite sensitive 
to the spatial
resolution. At relatively coarse resolutions, the low level of 
dissipation can give rise
to instability of the simulations. Moreover, these models add 
significantly to the
required computational effort. At suitable resolution, however, the 
predictions arising
from generalized similarity models are quite accurate.

An interesting subgrid model which follows the similarity approach to 
some degree and
avoids the costly additional filter-operations is the nonlinear 
gradient model, mentioned
earlier. This model can be derived from the Bardina scale-similarity 
model by using Taylor
expansions of the filtered velocity. One may arrive at
$\tau_{ij}= \recitwelve  \sum_k\Delta_k^2 (\pr_k \ub_i) (\pr_k \ub_j) 
+{\cal O}(\Delta^4)$.
The first term on the right-hand side is referred to as the 
`nonlinear gradient model' or
tensor-diffusivity model:
\be
m^{TD}_{ij}=\frac{1}{24} \sum_k\Delta_k^2 (\pr_k v_i) (\pr_k v_j)
\label{clark}
\en
Since this model is part of all three different \lesa models 
identified in the previous
subsection, we will analyze the dynamics and the instabilities
arising from this model in some more detail in the next subsection.

The three subgrid models, i.e., (\ref{smag}), (\ref{bardin}) and (\ref{clark})
constitute well-known examples in LES-literature, which represent 
basic dissipative and
reactive, or dispersive, properties of subgrid models for the 
turbulent stress tensor.
These basic similarity and eddy-viscosity models can be combined in 
mixed models using the
dynamic procedure, which provides a way of combining the two basic 
components of a mixed
model without introducing additional external ad hoc parameters.

We consider simple mixed models based on eddy-viscosity and similarity.
In these models the eddy-viscosity component reflects local
turbulence activities and the local value of the eddy-viscosity 
adapts itself to the
instantaneous flow. The dynamic procedure starts from Germano's 
identity (\ref{tradgerm}).
A common way to write Germano's identity is:
\be
T_{ij}-\widehat{\tau_{ij}}=R_{ij}
\label{iden}
\en
where
\bea
T_{ij} &=& \widehat{\ovl{ u_i u_j}}-
\widehat{\ovl{u}}_i \widehat{\ovl{ u}}_j \\
R_{ij} &=& {\widehat{(\ovl{ u}_i \ovl{u}_j)}}
-\widehat{\ovl{u}}_i \widehat{\ovl{u}}_j
\label{L}
\ena
Here, in addition to the basic LES-filter $\ovl{(\cdot)}$ of width 
${\ovl{\Delta}}$
a so-called `test'-filter $\widehat{(\cdot)}$ of width ${\widehat{\Delta}}$ is
introduced. Usually, this test-filter is wider than the LES-filter 
and the combined filter
${\widehat{\ovl{(\cdot)}}}$ has a width that follows from
${\widehat{\ovl{\Delta}}}^{2}={\widehat{\Delta}}^{2}+{\ovl{\Delta}}^{2}$. 
This relation
is exact for the composition of two Gaussian filters and can be shown 
to be `optimal' for
other filters such as the top-hat filter \cite{vreman_jfm}. The only 
external parameter
that needs to be specified in the dynamic procedure is the ratio
${\widehat{\Delta}}/{\ovl{\Delta}}$ which is commonly set equal to 
two. The terms at the
left-hand side of the Germano  identity (\ref{iden}) are the 
turbulent stress tensor on
the `combined' filter level ($T_{ij}$) and the turbulent stress 
tensor, filtered with the
test-filter ($\widehat{\tau_{ij}}$), respectively. Finally, $R_{ij}$ 
represents the
resolved stress tensor which can be explicitly calculated using the 
modeled LES fields.

The general procedure for obtaining `locally' optimal model parameters
in a mixed formulation starts by assuming a basic model $m_{ij}$ to 
approximate the
turbulent stress tensor $\tau_{ij}$, and a corresponding model 
$M_{ij}$ for $T_{ij}$.
We consider $m_{ij}$ to be of `mixed' type, i.e.,
\be
m_{ij}=a_{ij}+c b_{ij}
\en
where $a_{ij}$ and $b_{ij}$ are basic models. These basic models involve
operations on ${\bf{v}}$ only; $a_{ij}=a_{ij}({\bf{v}})$, 
$b_{ij}=b_{ij}({\bf{v}})$.
Furthermore, in standard mixed models, $c$ is a scalar 
coefficient-field which is to be
determined. The model $M_{ij}$ is represented as:
\be
M_{ij}=A_{ij}+C B_{ij}
\en
where $A_{ij}=a_{ij}({\widehat{{\bf{v}}}})$, 
$B_{ij}=b_{ij}({\widehat{{\bf{v}}}})$. It is
essential in this formulation that the coefficient $C$ corresponding 
to the composed
filter-level is well approximated by the coefficient $c$; i.e., we 
assume $C \approx c$.
Insertion in Germano's identity yields
$ M_{ij}+{\widehat{m_{ij}}}=R_{ij}$, or in more detail,
\be
\Big(A_{ij}+\widehat{a_{ij}}\Big) + c \Big[ B_{ij} + \widehat{b_{ij}} 
\Big]=R_{ij}
\en
where we have used the approximation ${\widehat{cb_{ij}}} \approx c 
{\widehat{b_{ij}}}$.
Introducing the short-hand notation ${\cal 
A}_{ij}=A_{ij}+\widehat{a_{ij}}$, ${\cal
B}_{ij}= B_{ij} + \widehat{b_{ij}}$, the coefficient $c$ is required 
to obey $c{\cal
B}_{ij}=R_{ij}-{\cal A}_{ij}$. This relation should hold for all 
tensor-components, which
of course is not possible for a scalar coefficient field
$c$. To resolve this situation we introduce an averaging operator 
$\langle f \rangle $
and define the `Germano-residual' by
\be
\varepsilon(c)= \langle \frac12 \{(R_{ij}-{\cal A}_{ij})-c{\cal 
B}_{ij}\}^{2} \rangle
\en
From this we obtain an optimality condition for $c$ from 
$\varepsilon'(c)=0$ and we can solve the local coefficient as
\be
c=\frac{\langle (R_{ij}-{\cal A}_{ij}){\cal B}_{ij}\rangle}{\langle 
{\cal B}_{ij}{\cal B}_{ij}\rangle}
\label{dyncoef}
\en
where we assumed $\langle c fg \rangle \approx c \langle f g 
\rangle$. The averaging
operator $\langle f \rangle$ is usually defined in terms of an integration over
homogeneous directions of the flow-domain. In the case of the mixing 
layer, considered
here, the averaging over the homogeneous streamwise and spanwise 
direction results in a
dynamic coefficient $c$ which is a function of the normal coordinate 
$x_{2}$ and time $t$.
In more complex flow-domains, averaging over homogeneous directions 
may no longer be
possible. Taking a running-average over time $t$ is then a viable 
alternative, as was
recently established,  e.g., for flow in a spatially developing mixing layer
\cite{debruin}.

As an example we consider the Smagorinsky model as the base model. 
The corresponding
models on the two filter-levels can be written as
\be
m^{D}_{ij}=- C_d {\ovl{\Delta}}^2 |\sigma({\bf v})|
\sigma_{ij}({\bf v})~~;~~
M^{D}_{ij}=- C_d {{\widehat{\ovl{\Delta}}}}^2
|\sigma(\widehat{\bf v})| \sigma_{ij}(\widehat{\bf v})
\label{german}
\en
The `optimal' $C_{d}$ follows from
$C_{d}=\langle R_{ij}{\cal B}_{ij} \rangle / \langle {\cal 
B}_{ij}{\cal B}_{ij}\rangle$.
In order to prevent numerical instability caused by negative values
of $C_d$, the model coefficient $C_d$ is artificially set to zero at
locations where the procedure would return negative values. 
Sometimes, in developing
flows, it is beneficial to also introduce a `ceiling'-value for 
$C_{d}$. This value should
be chosen such that once the flow is well-developed in time the 
actual limitation arising
from the ceiling-value is no longer restrictive \cite{debruin}.

The dynamic mixed model employs the sum of Bardina's similarity and Smagorinsky
eddy-viscosity model as the base model, i.e.,
\be
m^{DM}_{ij}= [L,S](v_{i},v_{j}) - C_d  {\ovl{\Delta}}^2 |S({\bf{v}})| 
S_{ij}({\bf{v}})
\label{dynmix}
\en
Likewise, a mixed nonlinear gradient model can be introduced by
\be
m^{DG}_{ij}= \frac{1}{24} {\ovl{\Delta}}^2 (\pr_k v_i) (\pr_k v_j)
- C_d  {\ovl{\Delta}}^2 |S({\bf{v}})| S_{ij}({\bf{v}})
\label{dynCla}
\en
The dynamic procedure has been used in a number of different flows. Compared to
predictions using only the constitutive base models, the dynamic 
procedure generally
enhances the accuracy and robustness. Moreover, it responds to the 
developing flow in such
a way that the eddy-viscosity is strongly reduced in laminar regions 
and near solid walls
\cite{wasistho}. This avoids specific modeling of transitional 
regions and near-wall
phenomena, provided the resolution is sufficient. At even coarser 
resolution one may have
to resort to specific models for transition and walls. We will not 
enter into this
problem. Rather, we will focus on the properties of the nonlinear 
gradient model in the
next subsection.

\subsection{Analysis of instabilities of the nonlinear gradient model 
in one dimension}
\label{sec2-4}

 From the discussion of the previous two subsections, it would appear 
that the nonlinear
gradient subgrid model would be very well suited to parameterize the 
dynamic effects of
the small scales in a turbulent flow. This model is part of the full 
\lesa model and it
also emerges as a Taylor expansion of the Bardina similarity model. 
In this subsection we
will analyse the nonlinear gradient model in the context of the 
one-dimensional Burgers
equation and show that this model gives rise to very strong 
instabilities. Apparently,
some features appear to be missing in the pure nonlinear gradient 
model. In subsequent
sections we will show in what way the explicit filtering and the 
other terms in the \lesa
model, or dynamic eddy-viscosity regularization, alter this peculiar 
behavior of the
nonlinear gradient model.

We will analyse the nature of the instability of the pure gradient 
model for the
one-dimensional Burgers equation \cite{vreman_tcfd}. The linear 
stability of a sinusoidal
profile will be investigated. If a flow is linearly
unstable then it is nonlinearly unstable to arbitrarily small initial
disturbances. The linear analysis thus provides some information on
the nonlinear equation.

The Burgers equation with gradient subgrid-model is written as:
\be
\pr_t u+\half \pr_x (u^2)-\nu \pr_x^2 u =-\half \eta \pr_x(\pr_x u)^2+f(x)
\label{Burgers1}
\en
The parameter $\eta=\Delta^2/12$. The following analysis shows that 
smooth solutions of
equation (\ref{Burgers1}) can be extremely sensitive to small
perturbations, leading to severe instabilities.
In particular, we consider the linear stability of a $2\pi$-periodic,
stationary solution, $U(x,t)=\sin(x)$ on the
domain $[0,2\pi]$ with periodic boundary conditions. The forcing
function $f$ is determined by the requirement
that $U$ is a solution of equation (\ref{Burgers1}).
We substitute a superposition of $U$ and a perturbation $w$,
\be
u(x,t)=U(x)+w(x,t)
\en
into equation (\ref{Burgers1}) and linearize around $U$, omitting higher
order terms in $w$:
\be
\pr_t w+(1-\eta)\sin(x)~\pr_x w+ (w+\eta \pr_x^2 w)
\cos(x)=\nu
\pr_x^2 w
\label{veqn}
\en
We use a Fourier expansion for $w$ written as $w=\sum \alpha_k (t) e^{ikx}$.
After substitution of this series
into equation (\ref{veqn}) we obtain an infinite
system of ordinary differential equations for the Fourier
coefficients $\alpha_k$:
\be
\dot{\alpha}_k = \half k (\eta k -\eta -1)
\alpha_{k-1} -k^2 \nu \alpha_k +
\half k(\eta k +\eta +1)
\alpha_{k+1}
\label{system}
\en

To understand
the nature of the nonlinear gradient model for the Burgers equation,
we first analyse system (\ref{system}) assuming $\nu=0$. Instead of
the infinite system, we consider a sequence of finite dimensional
systems,
\be
\dot{{\bf z}}_n=M_n {\bf z}_n
\en
where ${\bf z}_n$ is a vector containing the $2n+1$ Fourier
coefficients $\alpha_{-n}...\alpha_{n}$
and $M_n$ is a $(2n+1)\times (2n+1)$ tri-diagonal matrix:
\be
{\bf z}_n=\left[ \begin{array}{l}
  \alpha_{-n} \\ . \\ . \\ \alpha_{-1} \\
\alpha_0 \\ \alpha_{1} \\ . \\ . \\ \alpha_n \end{array} \right],
~~~~~~~M_n=\left[ \begin{array}{ccccccccc}
  0 & l_n &     &   & & & &         &  \\
  r_n & . & .   &   & & & &         & \\
    & . &  .  &  l_2 &  & & &          & \\
    & & r_2 &  0  & l_1 &  & &       & \\
    &    &  &  0  &  0  & 0  & & &  \\
    &    &  &    &  l_1 & 0 & r_2 &&      \\
    &     &    &      & &   l_2 & . & .  & \\
    &     &     &   &  & &  . & .      & r_n \\
    &     &     &   & & & & l_n & 0
\end{array} \right]
\en
with
\be
l_k = \half k (\eta k -\eta -1) \label{lk}~~~;~~~
r_k = \half (k-1) (\eta k +1 ) \label{rk}
\en

The eigenvalues of $M_n$ determine the stability of the problem. The
system is unstable if
the maximum of the real parts of the eigenvalues is positive.
We denote the eigenvalues of
$M_n$ by $\lambda_j$ and introduce $\lambda_{max}$ such that
\be
|\lambda_{max}|=\max_{j}|\lambda_j|
\en
This eigenvalue problem can be shown to have the following asymptotic 
properties (for a
detailed proof see \cite{vreman_tcfd}):
\bea
1. & &
\mbox{if}~~ \lambda ~~\mbox{is an eigenvalue then}~ -\lambda~~ \mbox{is an
eigenvalue} \label{point1} \\
2. & & |\lambda_{max}|\sim \eta n^2 \label{point2} \\
3. & & |\mbox{Im}(\lambda_{max})|\leq n-1 \label{point3}
\ena
The first point implies that $\lambda_{max}$ can be chosen such that
$\mbox{Re}(\lambda_{max})\geq 0$.
Hence, the combination of these three properties
yields the asymptotic behavior of the maximum of the real
parts of the eigenvalues:
\be
\mbox{Re}(\lambda_{max})\sim \eta n^2
\label{est3}
\en
This shows that the inviscid system is linearly unstable and
that the largest real part of the eigenvalues is asymptotically
proportional to $n^2$, where $n$ is the number of Fourier modes taken into
account.

It should be observed that the instability is severe, since the system
is not only unstable, but the
growth rate of the instability is infinitely large as $n \rightarrow
\infty$.
The instability is
fully due to the incorporation of the gradient model, since all eigenvalues
of the matrix $M_n$ are purely imaginary in case the inviscid
Burgers equation without subgrid-model is considered ($\eta=0$).
In numerical
simulations the instability will grow with a finite speed, since
then the number of Fourier modes is limited by the finite grid.
Moreover, expression (\ref{est3}) illustrates that
grid-refinement (with $\eta$ kept constant), which corresponds to a
larger $n$, will not stabilize the system, but rather enhance the
instability. The growth rate of
the instability of the one-dimensional problem
can be expressed in terms of $\Delta$ and the grid-spacing $h$:
$\eta n^2 \sim (\Delta /h)^2$. Consequently, the instability is not
enhanced if the ratio between $\Delta$ and $h$ is kept constant.

Finally, we will consider the more complicated case $\nu \neq 0$. The
linear system in equation (\ref{system}) now gives rise to matrices
$M_n$ which have a negative principal diagonal.
It is known that for every fixed value of $n$ there exists an
eigenvalue arbitrarily close to the eigenvalue of the inviscid system
($\lambda_{max}$)
if $\nu$ is sufficiently small \cite{chatelin}.
Hence for small values
of $\nu$ the viscous system for finite $n$ is still linearly unstable.
The matrix $M_n$ is
strictly diagonally dominant if $\nu > \eta+1 $,
while all rows except $n$ and $n+2$ are already diagonally dominant if
$\nu > \eta$. If the matrix is diagonally dominant, the real parts of
all eigenvalues are negative and, consequently, the system is stable.
This indicates that stability can be achieved by a sufficiently
large viscosity, which does not depend on $n$, but only on $\eta$. 
Thus, if the gradient
model is supplemented with an adequate eddy-viscosity the instability 
will be removed as
is the case with a dynamic mixed model involving the gradient model.

\section{Numerical simulations of a turbulent mixing layer}
\label{sim}

In this section we first present the numerical methods used to
solve the DNS and LES equations (subsection~\ref{nummeth}). We 
illustrate the accuracy of
these methods for turbulent flow in a mixing layer in subsection~\ref{mxl}.

\subsection{Time-integration and spatial discretization}
\label{nummeth}

The Navier-Stokes or modeled LES equations are discretized using the 
so-called method of
lines. We consider the compressible formulation and perform 
simulations at a low
convective Mach number which was shown to provide essentially incompressible
flow-dynamics. The method of lines allows to treat the spatial and temporal
discretization separately and gives rise to a large number of 
ordinary differential
equations for the unknowns on a computational grid.

We write the Navier-Stokes or LES equations concisely as 
$\pr_{t}{\cal U}= {\cal F}({\cal
U})$ where ${\cal U}$ denotes the state-vector containing,  e.g., 
velocity and pressure,
and ${\cal F}$ is the total flux, composed of the convective, the viscous, and
possibly the subgrid fluxes. The operator ${\cal F}$ contains first 
and second order
partial derivatives with respect to the spatial coordinates $x_{j}$. 
The equations are
discretized on a uniform rectangular grid and the grid size in the 
$x_j$-direction is
denoted by $h_j$. If we adopt a specific spatial discretization 
around a grid point ${\bf
x}_{ijk}$, the operator ${\cal F}({\cal U})$ is approximated in a 
consistent manner by an
algebraic expression
$F_{ijk}(\{U_{\alpha \beta \gamma}\})$ where $\{ U_{\alpha \beta \gamma} \}$
denotes the state vectors in all the grid-points, labeled by 
$\alpha,~\beta,~\gamma$.
Usually, only neighboring grid points around $(i,~j,~k)$ appear 
explicitly in $F_{ijk}$,
e.g., in case finite difference or finite volume discretizations are 
considered. After
applying the method of lines, the governing equations yield
\be
{\rm d}_{t}U_{ijk}(t)=F_{ijk}(\{U_{\alpha \beta \gamma} 
\})~~~;~~U_{ijk}(0)=U_{ijk}^{(0)}
\en
where $U_{ijk}^{(0)}$ represents the initial condition. Hence, in 
order to specify the
numerical treatment, apart from the initial and boundary conditions, 
the spatial
discretization which gives rise to $F_{ijk}$ and the temporal 
integration need to be
specified. We next introduce these separately.

The time stepping method which we adopt is an explicit
four-stage compact-storage
Runge-Kutta method. When we consider the scalar differential
equation $du/dt=f(u)$, this Runge-Kutta method performs within
one time step of size $\delta t$
\be
u^{(j)}=u^{(0)}+\beta_j \delta t f(u^{(j-1)})~~~(j=1,2,3,4)
\en
with
$u^{(0)}=u(t)$ and $u(t+\delta t)=u^{(4)}$. With the coefficients
$\beta_1=1/4$,
$\beta_2=1/3$, $\beta_3=1/2$ and $\beta_4=1$ this yields
a second-order accurate time integration
method \cite{jameson}. The time step is determined by the
stability restriction of the numerical scheme. It depends on the grid-size $h$
and the eigenvalues of the flux Jacobi matrix of the numerical flux $f$. In a short-hand
notation one may write $\delta t = {\rm CFL}~ h / |\lambda_{max}|$ where $|\lambda_{max}|$ denotes
the eigenvalue of the flux Jacobi matrix with maximal size, and ${\rm CFL}$ denotes the Courant-Friedrichs-Levy-number
which depends on the specific choice of explicit time integration method.
For the present four-stage Runge-Kutta method a maximum CFL number of 2.4 can be established using a Von Neumann
stability analysis. In the actual simulations we use ${\rm CFL}=1.5$, which is suitable for both DNS and LES,
irrespective of the specific subgrid model used.

In order to specify the spatial discretization we distinguish between 
the treatment of
the convective and the viscous fluxes. We will only specify the 
numerical approximation of
the $\pr_1$-operator; the $\pr_2$ and $\pr_3$-operators are treated 
analogously.
Subgrid-terms are discretized with the same method as the viscous 
terms. Throughout we will use a second order method for the
viscous fluxes and both a second order, and a fourth order accurate 
method for the
convective fluxes. All these methods are constructed from (a 
combination of) first order
numerical derivative operators $D_{j}$.

The second-order method that we consider is a finite volume method
\cite{kuerten_fv}.  The discretization of the convective terms is the
cell vertex trapezoidal rule, which is a weighted second-order central
difference. In vertex $(i,j,k)$ the corresponding operator is denoted
by $D_1$ and for the approximation of $\pr_{1}f$ it is defined as
\bea
(D_1 f)_{i,j,k} &=& (s_{i+1,j,k}-s_{i-1,j,k})/(2h_1) \label{D1} \\
\mbox{with}~~~~ s_{i,j,k} & =& (g_{i,j-1,k}+2g_{i,j,k}+g_{i,j+1,k})/4
\nonumber \\
\mbox{and}~~~~ g_{i,j,k} & =& (f_{i,j,k-1}+2f_{i,j,k}+f_{i,j,k+1})/4
\nonumber
\ena
The viscous terms contain second-order derivatives which are treated 
by a consecutive application of two
first order numerical derivatives. This requires for example that the 
gradient of the velocity
is calculated in centers of grid-cells. In
center $(i+\half,j+\half,k+\half)$ the corresponding discretization 
$D_2 f$ has the form
\bea
(D_2 f)_{i+\half,j+\half,k+\half} &=& (s_{i+1,j+\half,k+\half}
-s_{i,j+\half,k+\half})/h_1 \\
\mbox{with}~~~~ s_{i,j+\half,k+\half} & =&
(f_{i,j,k}+
f_{i,j+1,k}+f_{i,j,k+1}+f_{i,j+1,k+1})/4
\nonumber
\ena
The second derivative is subsequently calculated with operator $D_{1}$; thus we
approximate,  e.g., $\pr_{11}(f)_{ijk} \approx D_{1}(D_{2}(f))_{ijk}$.

The combination of $D_{1}$ and $D_{2}$ is robust with respect to 
odd-even decoupling but it is only second order accurate.
In a similar manner we may construct a fourth-order accurate method. 
The corresponding expression for $D_3 f$ has the following form:
\bea
(D_3 f)_{i,j,k} \! \! \! &=& \! \! \!
(-s_{i+2,j,k}+8s_{i+1,j,k}-8s_{i-1,j,k}+s_{i-2,j,k})/(12h_1) \label{D3} \\
\mbox{with}~~~~ s_{i,j,k}  \! \! \! & =&  \! \! \! (-g_{i,j-2,k}+4g_{i,j-1,k}+
10g_{i,j,k}+4g_{i,j+1,k}-g_{i,j+2,k})/16
\nonumber \\
\mbox{and}~~~~ g_{i,j,k}  \! \! \! & =& \! \! \! (-f_{i,j,k-2}+4f_{i,j,k-1}+
10f_{i,j,k}+4f_{i,j,k+1}-f_{i,j,k+2})/16
\nonumber
\ena
This scheme is conservative, since it is a weighted
central difference. The coefficients in the definition for
$g_{i,j,k}$ are chosen such that $g_{i,j,k}$ is a fourth order accurate
approximation to $f_{i,j,k}$ and $\pi$-waves in the $x_3$-direction
give no contributions to $g_{i,j,k}$. The definition for $s_{i,j,k}$
has the same properties with respect to the $x_2$-direction.
For convenience, we will refer to a combination of $D_{3}$ for the 
convective, and $D_{1},~D_{2}$ for the viscous fluxes
as fourth-order methods, but we remark that the
formal spatial accuracy of the scheme is only second-order due to the
treatment of the viscous terms.

\subsection{The turbulent mixing layer}
\label{mxl}

The flow in a temporally developing turbulent mixing layer is well 
documented in literature (e.g. \cite{vreman_jfm}),
and will be considered here
to test the \lesa modeling approach. In this section we review the 
scenario of the development of the flow that is considered
and sketch the type of predictions that can be obtained by 
traditional LES using the dynamic model. This serves as a point of
reference for the next section.

We simulate the compressible
three-dimensional temporal mixing layer and use a convective Mach 
number $M=0.2$
and a Reynolds number based
on upper stream velocity and half the initial vorticity thickness of 50.
The governing equations are solved
in a cubic geometry of side $\l=59$.
Periodic boundary conditions are imposed in the streamwise ($x_{1}$) and
spanwise ($x_{3}$) direction, while in the normal ($x_{2}$) direction 
the boundaries are
free-slip walls. The initial condition is formed by mean profiles
corresponding to constant pressure $p=1/(\gamma M^{2})$ where $\gamma=1.4$
is the adiabatic gas constant, $u_{1}=\tanh(x_{2})$ for the streamwise
velocity component, $u_{2}=u_{3}=0$ and a temperature profile given
by the Busemann-Crocco law. Superimposed on the mean profile
are two- and three-dimensional
perturbation modes obtained from linear stability theory. Further
details may be found in \cite{vreman_jem}.

\begin{figure}[htb]

\centering{
{\psfig{figure=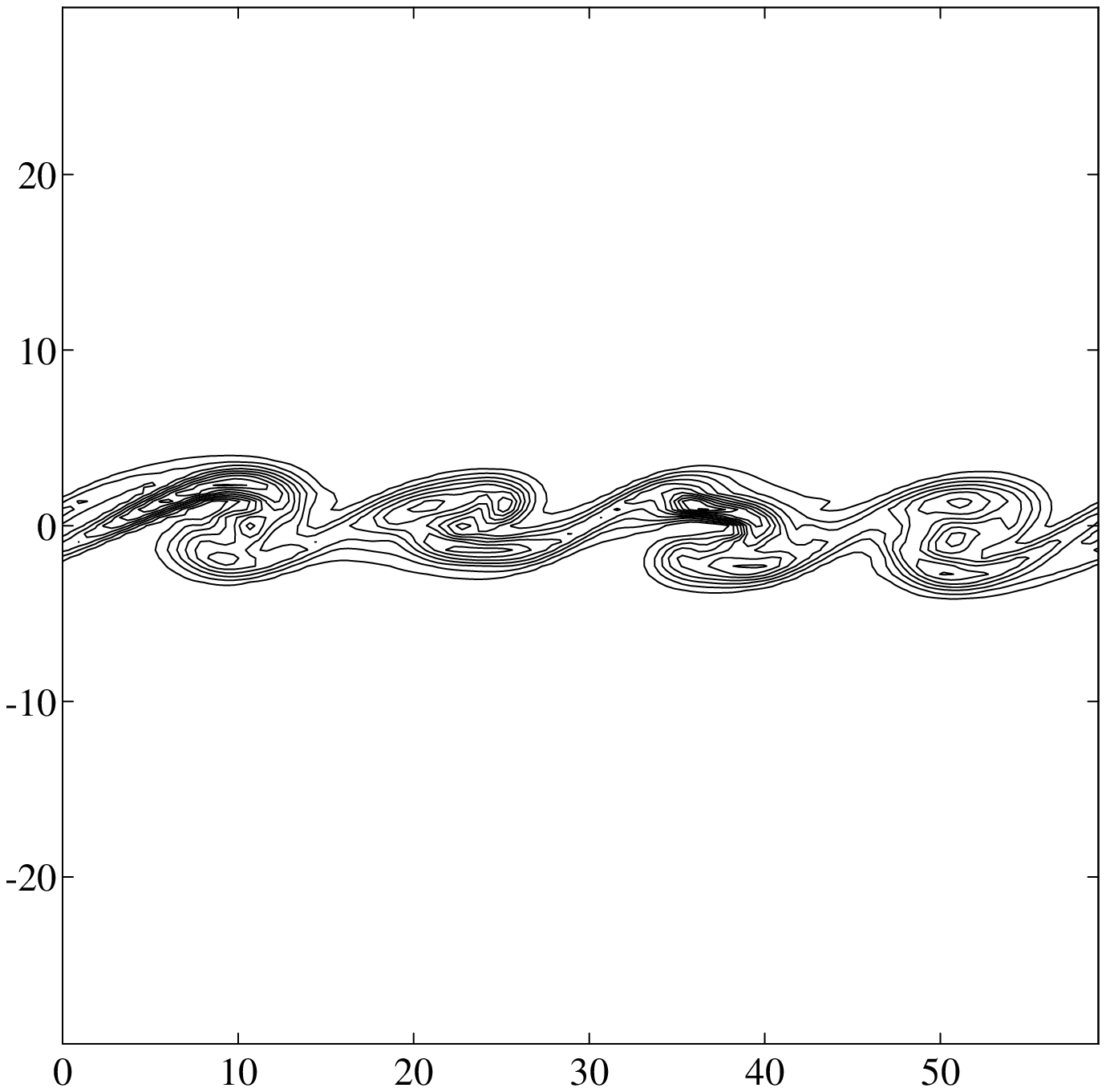,width=0.32\textwidth}}
{\psfig{figure=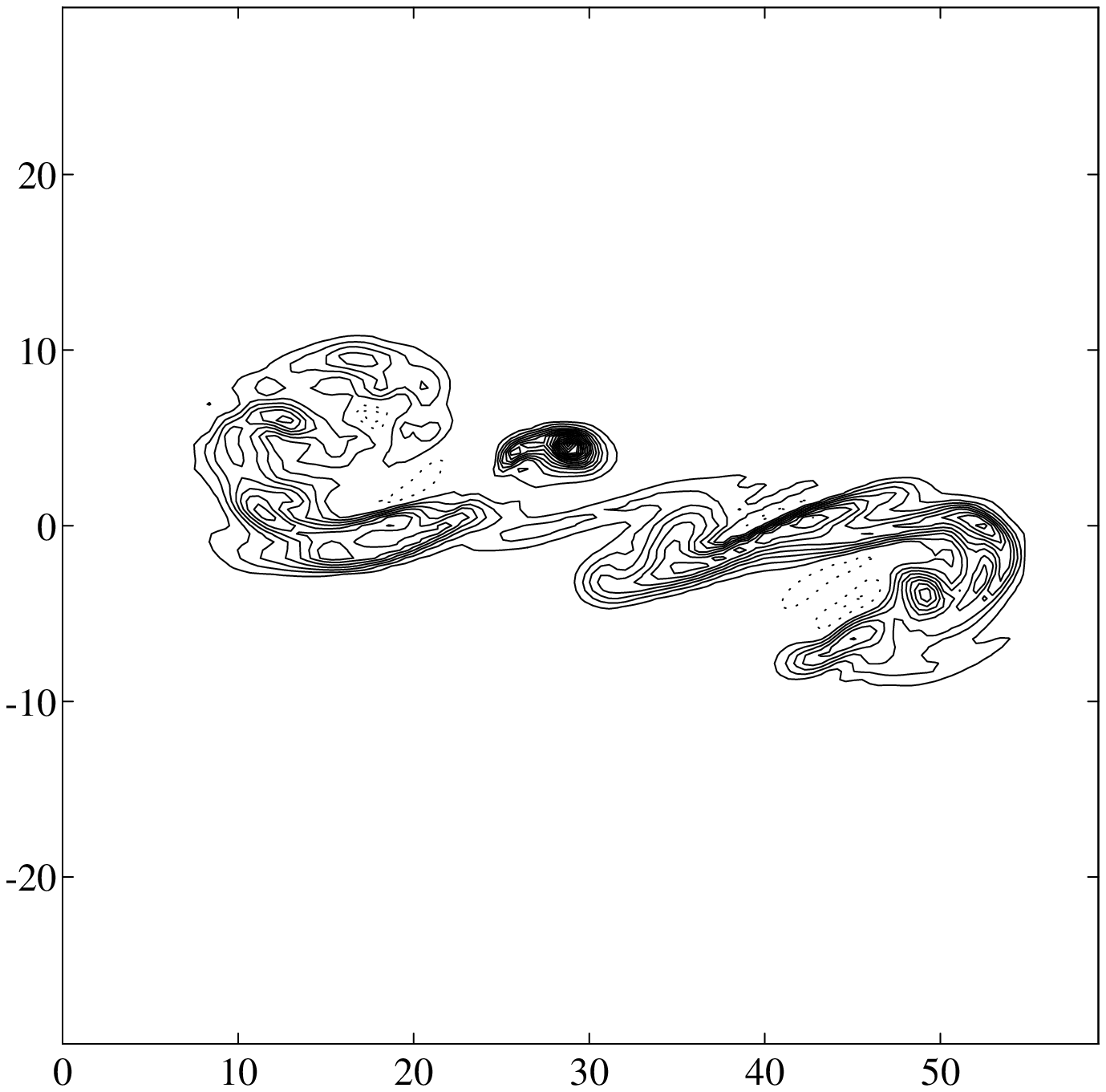,width=0.32\textwidth}}
{\psfig{figure=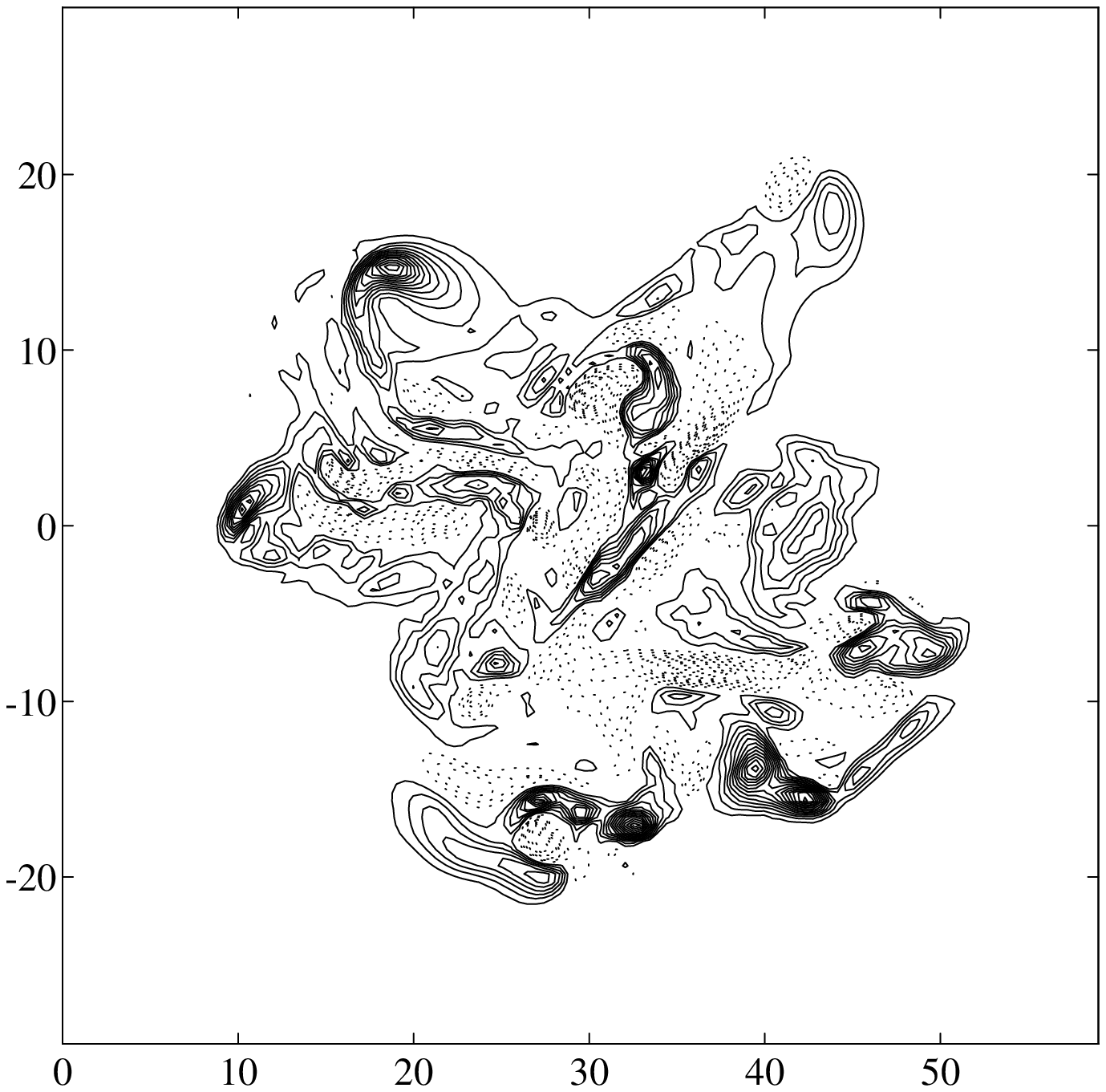,width=0.32\textwidth}}
}

\caption{\small{ Results from a DNS using $192^3$ points.}
  Contours of spanwise vorticity for the plane $x_3=3L/4$
at $t=20$, $t=40$ and $t=80$ from left to right. Solid and dotted contours
indicate negative and positive vorticity respectively. The contour
increment is 0.1.}
\label{mxldns}
\end{figure}

The DNS is conducted on a uniform grid
with $192^3$ cells using the fourth order spatial discretization method.
Visualization of the DNS data
demonstrates the roll-up of the fundamental instability and successive
pairings (figure~\ref{mxldns}). Four
rollers with mainly negative spanwise vorticity are observed at
$t=20$. After the first pairing ($t=40$) the flow has become highly
three-dimensional. Another pairing ($t=80$), yields a single roller in
which the flow exhibits a complex structure.

\begin{figure}[htb]

\centering{
{\psfig{figure=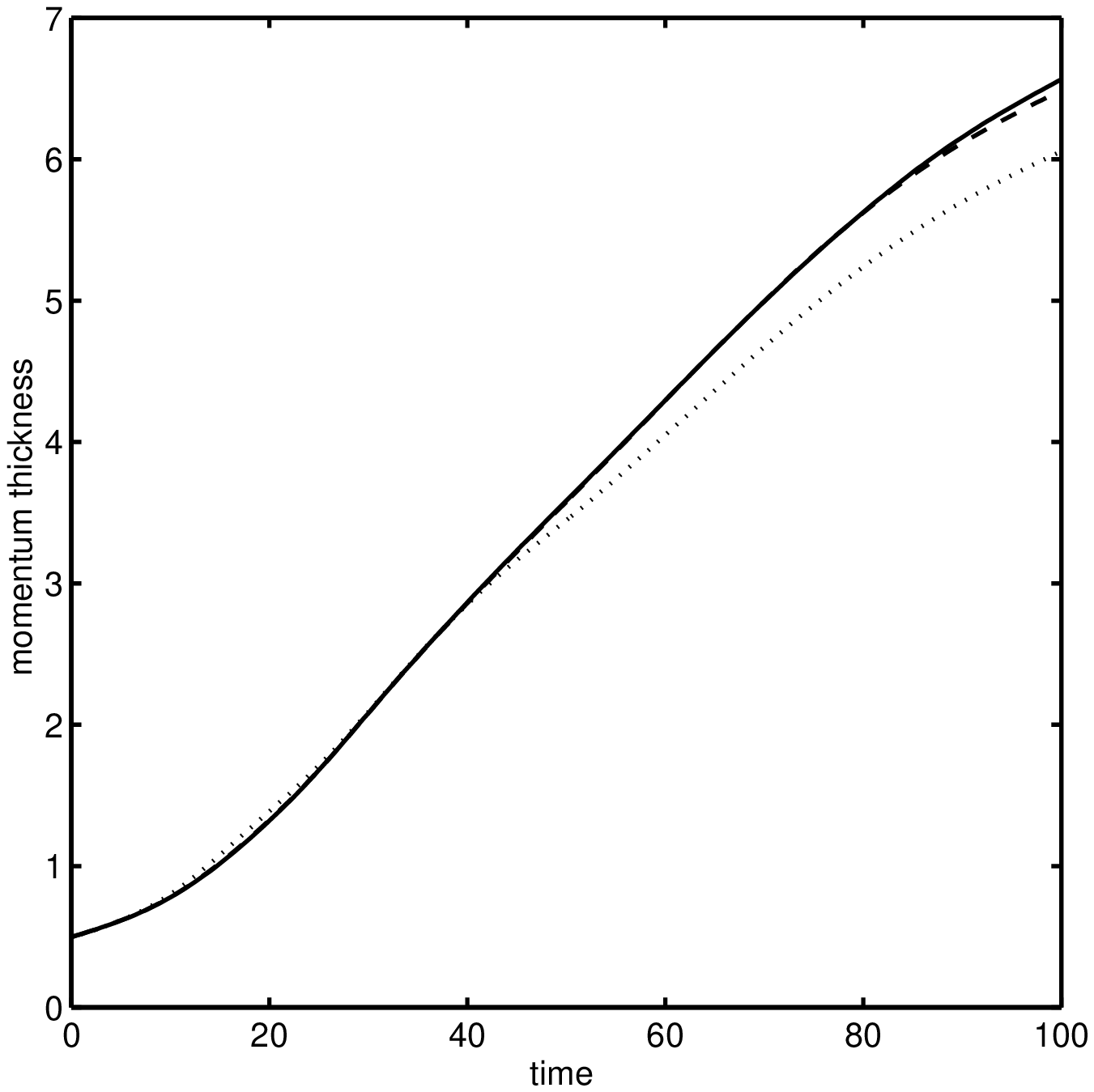,width=0.4\textwidth}}
{\psfig{figure=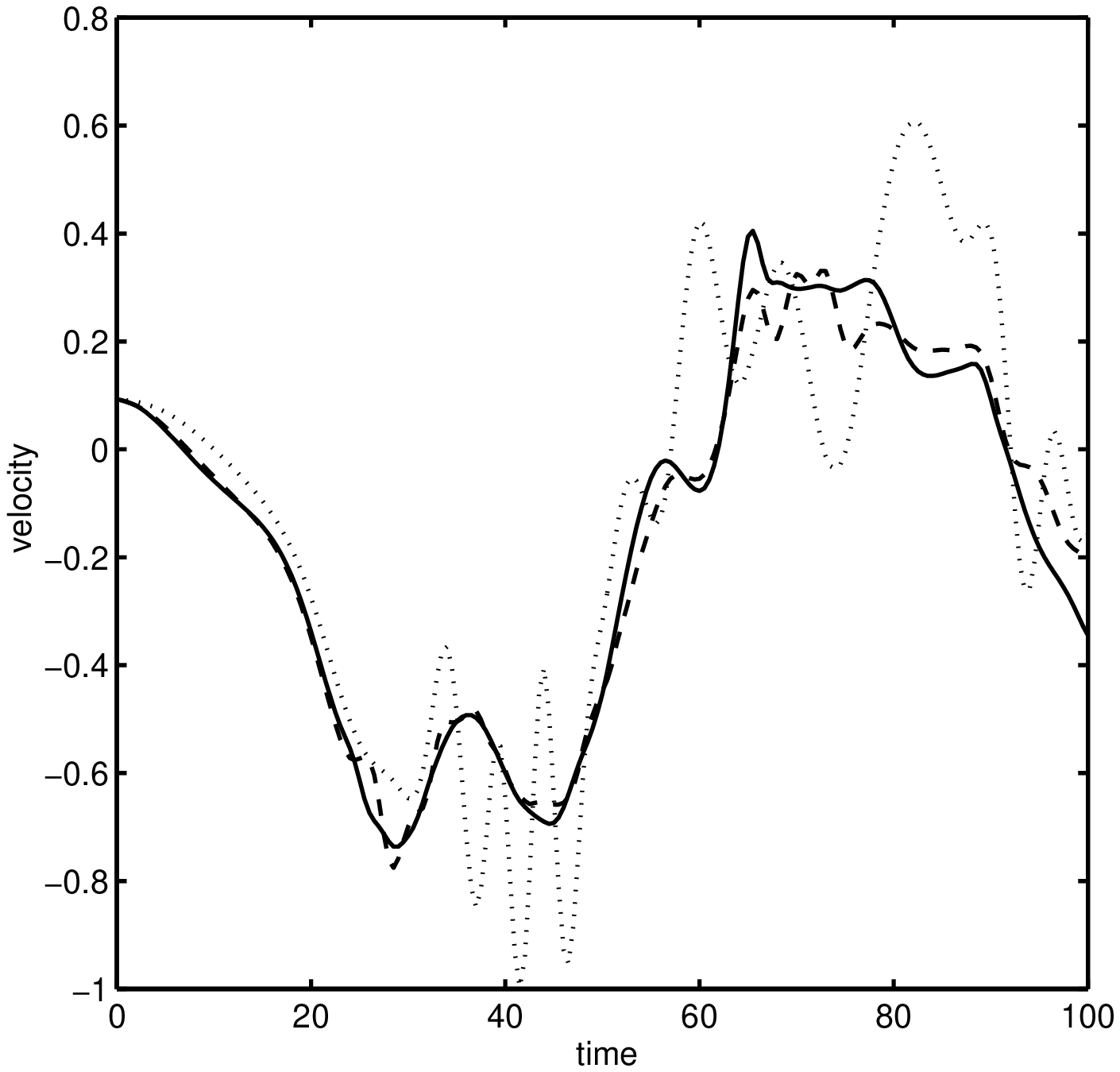,width=0.42\textwidth}}
}

\caption{\small Evolution of the momentum thickness (left) and
$u_3$ at $(\quarter L,0,\half L)$ (right) obtained from
simulations which do not involve any subgrid model and employ a
sequence of grids: $64^3$ (dotted), $128^3$ (dashed) and $192^3$ (solid).}
\label{dnsresolution}
\end{figure}

The accuracy of the simulation with $192^3$ cells is satisfactory as 
is inferred from
coarser grid computations on $64^3$ and $128^3$ cells.
The evolution of the momentum thickness
\be
\delta(t)=\frac14 \int_{-L/2}^{L/2} (1-\langle u_1 \rangle)(\langle 
u_1 \rangle +1)dx_2
\en
and an instantaneous velocity
component at the center of the shear layer are shown in 
figure~\ref{dnsresolution}.
The $64^3$-simulation is inadequate for the prediction of the
local instantaneous solution, but the momentum thickness appears 
quite reasonable.

\begin{figure}[htb]

\centering{
{\psfig{figure=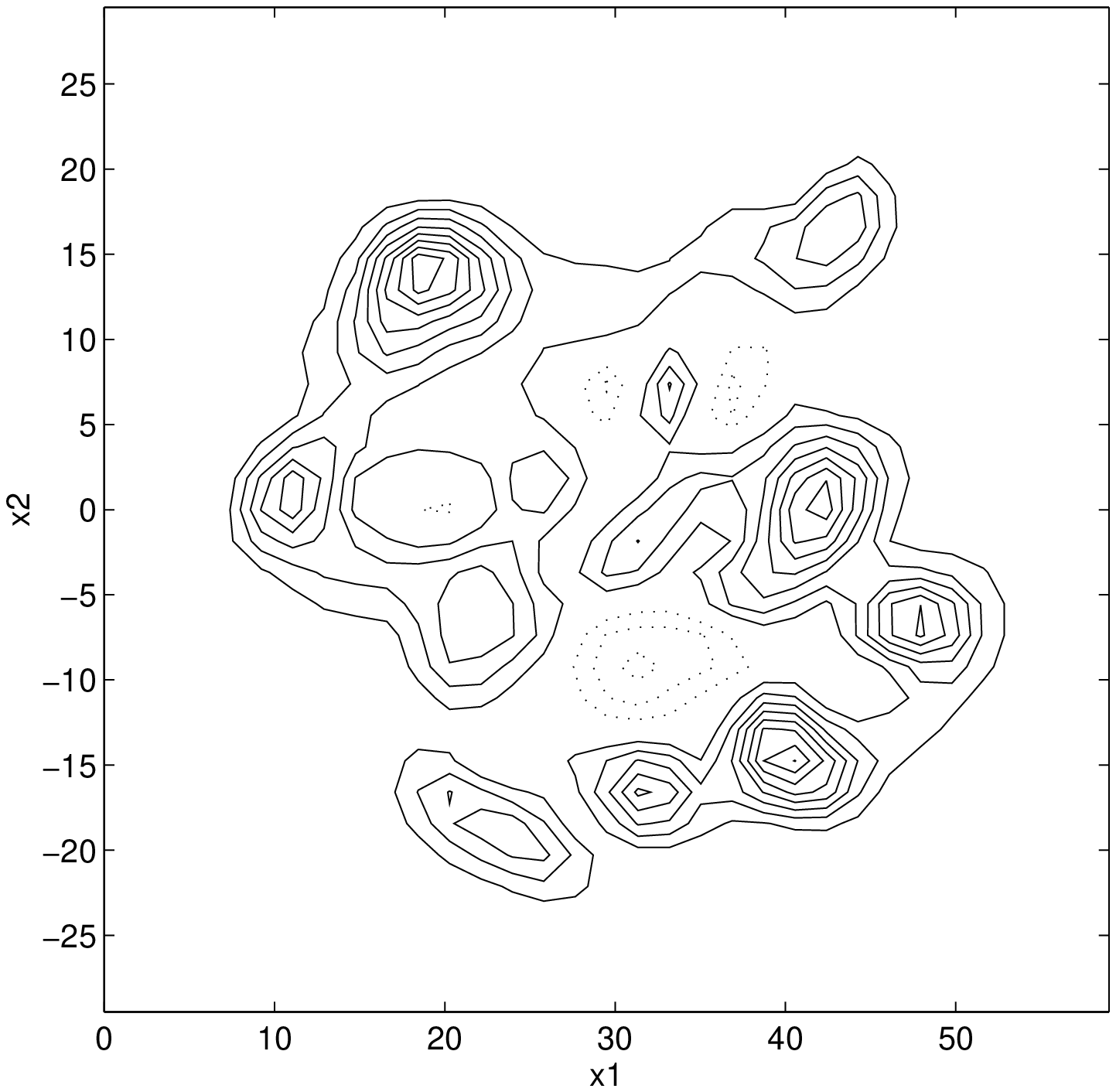,width=0.4\textwidth}}
{\psfig{figure=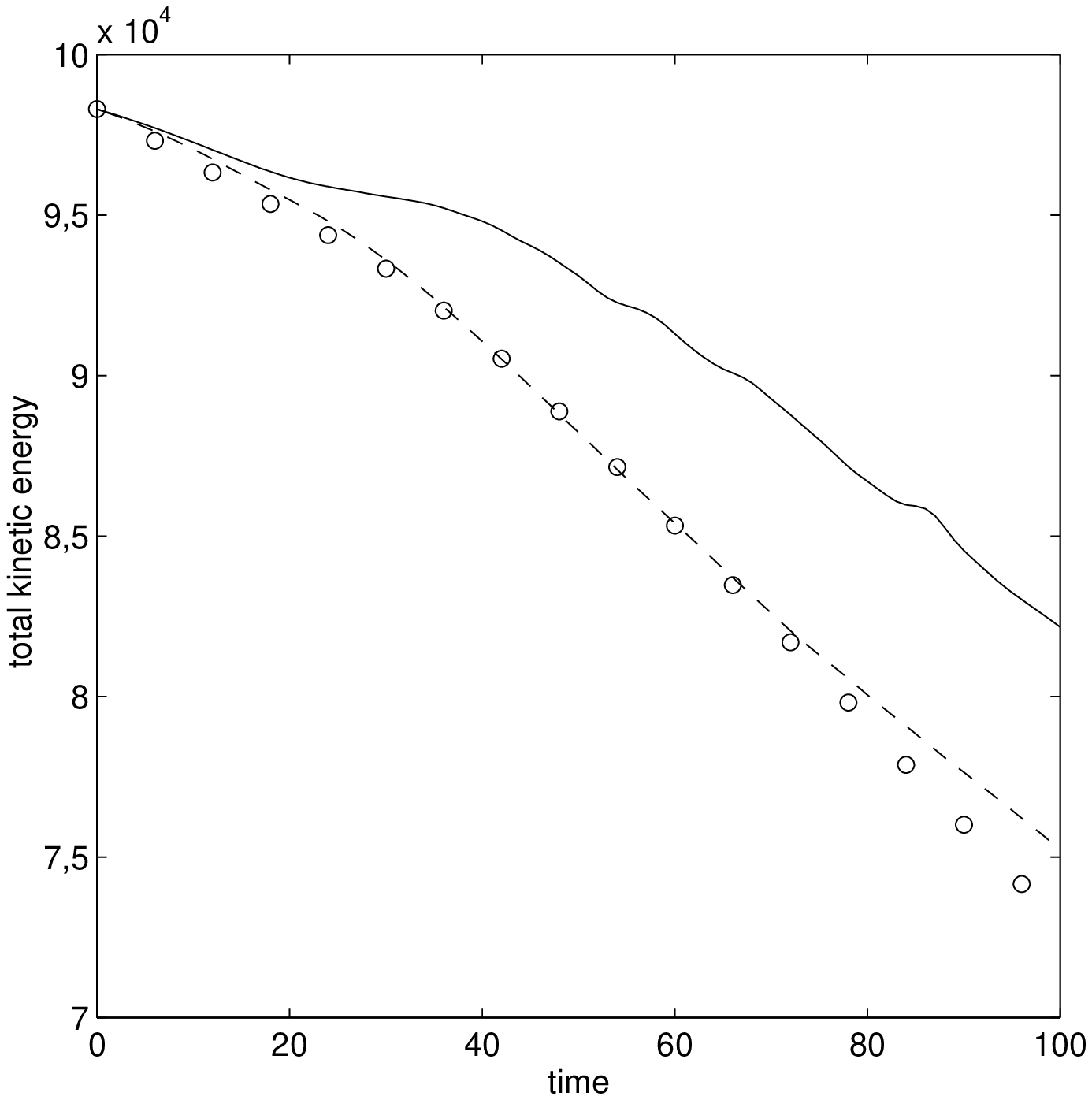,width=0.4\textwidth}}
}

\caption{\small { Left:} Contour-lines of the $z$-component of the
vorticity.
The effect of spatially filtering
the DNS solution at $t=80$ in figure~\ref{mxldns} with a top-hat 
filter and filter-width
$\Delta=L/16$.
{ Right:} prediction of the kinetic energy with
the dynamic eddy-viscosity model (dashed) compared with the filtered 
DNS results
(markers) and a simulation on the coarse LES grid ($32^{3}$) without 
a model (solid).}

\label{lesfig}

\end{figure}

To illustrate the effect that filtering has on a well-developed DNS 
solution,  vorticity
contours { for $\Delta=L/16$} are shown in figure~\ref{lesfig}.
Comparing this with the corresponding DNS results in 
figure~\ref{mxldns} allows one to
appreciate the strong smoothing effect that filtering has on the solution.
On the right in figure~\ref{lesfig} we included the decay of the 
resolved turbulent
kinetic energy, defined as
\begin{equation}
E= \frac12 \int_{\Omega} (\overline{u}_{1}  ^{2}
+ \overline{u}_{2} ^{2} + \overline{u}_{3} ^{2}) ~d {\bf{x}}
\label{kinenergy}
\end{equation}
We observe that the dynamic eddy-viscosity model generates quite a 
correction of the
`no-model coarse grid simulation'. Other models were also considered 
in \cite{vreman_jfm},
such as Smagorinsky's model, Bardina's scale-similarity model and 
dynamic mixed models.
Roughly speaking, the use of Bardina's model leads to flow 
predictions which contain
somewhat too many small scale features whereas the Smagorinsky model,
with eddy-coefficient $C_{S}=0.17$ prevents the flow from developing beyond the
transitional stage due to excessive dissipation in the early stages 
of the {evolution}.
Finally, the dynamic mixed models were all shown to perform about 
equally well and
provide accurate predictions.

\section{\lesa of a mixing layer}
\label{lesmxl}

In this section we will consider LES using the \lesa model. Above, in
section~\ref{sec2-2}, we introduced this model and identified three distinct
contributions; in fact, the \lesa model contains the explicitly 
filtered nonlinear
gradient model ($m_{ij}^{NG}$), the Leray model ($m_{ij}^{L}$) and 
the complete \lesa
model ($m_{ij}^{\alpha}$). These are defined as
\begin{eqnarray}
m_{ij}^{NG}&=&\frac{\Delta^{2}}{24}
\Big( \ovl{\pr_{k}v_{i}~\pr_{k}v_{j}} \Big) \equiv \ovl{A_{ij}}
\\
m_{ij}^{L}&=&\frac{\Delta^{2}}{24}
\Big( \ovl{\pr_{k}v_{i}~\pr_{k}v_{j}}+\ovl{\pr_{k}v_{i}~ \pr_{j}v_{k}}
  \Big) \equiv \ovl{A_{ij}}+\ovl{B_{ij}}
\\
m_{ij}^{\alpha}&=&\frac{\Delta^{2}}{24}
\Big( \ovl{\pr_{k}v_{i}~\pr_{k}v_{j}}+\ovl{\pr_{k}v_{i}~ \pr_{j}v_{k}}
-\ovl{\pr_{i}v_{k}~\pr_{j}v_{k}} \Big)
\nonumber \\
&\equiv& \ovl{A_{ij}}+\ovl{B_{ij}}-\ovl{C_{ij}}
\end{eqnarray}

First we will consider reference LES using these models and compare 
predictions with
those obtained with dynamic subgrid models (subsection \ref{refles}). 
Then we focus our
attention on the resolved kinetic energy dynamics in subsection 
\ref{energy}. Finally, in
subsection \ref{gridindependent} we consider (nearly) grid-independent
\lesa predictions which arise when refining the grid while keeping 
$\Delta$ constant.

\subsection{Reference LES of the mixing layer}
\label{refles}

In order to create a point of reference, we consider LES defined on a 
resolution of
$32^{3}$ grid-points. This choice represents a significant saving 
compared to the full DNS
and places a considerable importance on the subgrid fluxes. This resolution
was used previously in a comparative study of subgrid models in 
\cite{vreman_jfm}.

The simulations will be illustrated by considering the evolution of 
the resolved kinetic
energy $E(t)$, defined in (\ref{kinenergy}). In addition, we consider 
the momentum
thickness $\delta(t)$, based on filtered variables which quantifies 
the spreading of the
mean velocity profile.  We also investigate the
Reynolds-stress profiles $\langle w_{1}w_{2} \rangle$
defined with respect to the fluctuation $w_{i}=v_{i}-\langle v_{i} \rangle$.
Finally, we incorporate the streamwise kinetic energy spectrum in the 
turbulent regime  at
$t=80$. In this way a number of essentially
different quantities (mean, local, plane averaged) are included in the
comparisons in order to assess various aspects of the quality of the models.

For all simulations we will use a LES-filter-width $\Delta=L/16$. On 
the $32^{3}$ grid
this implies that $\Delta/h=2$, i.e., two grid-intervals cover the 
filter-width. Moreover,
unless explicitly stated otherwise, the explicit filter used in the 
definition of the
\lesa subgrid models will have the same width as the LES-filter, 
i.e., $\kappa=1$. The
filtering is done using the top-hat filter and we adopt the 
trapezoidal rule to perform
the numerical integrations. The simulations that will be presented in 
the following
subsections correspond to a slightly different initial condition than used in
section~\ref{mxl}. The differences are fairly small, but still 
prevent a direct comparison
with the filtered DNS results presented in section~\ref{mxl}.

\subsubsection{Explicit filtering is essential}

  The proposed subgrid models in the $\alpha$ framework each contain 
the nonlinear gradient
model and also involve an explicit filtering. As analyzed in 
section~\ref{sec2-4},
the nonlinear gradient model, without explicit filtering gives rise 
to instabilities.
These instabilities manifest themselves,  e.g., by an increase in the 
resolved kinetic
energy, instead of the monotonous decrease that is characteristic of 
this relaxing shear
layer, cf. figure~\ref{lesfig}.

\begin{figure}[htb]

\centering{
\psfig{figure=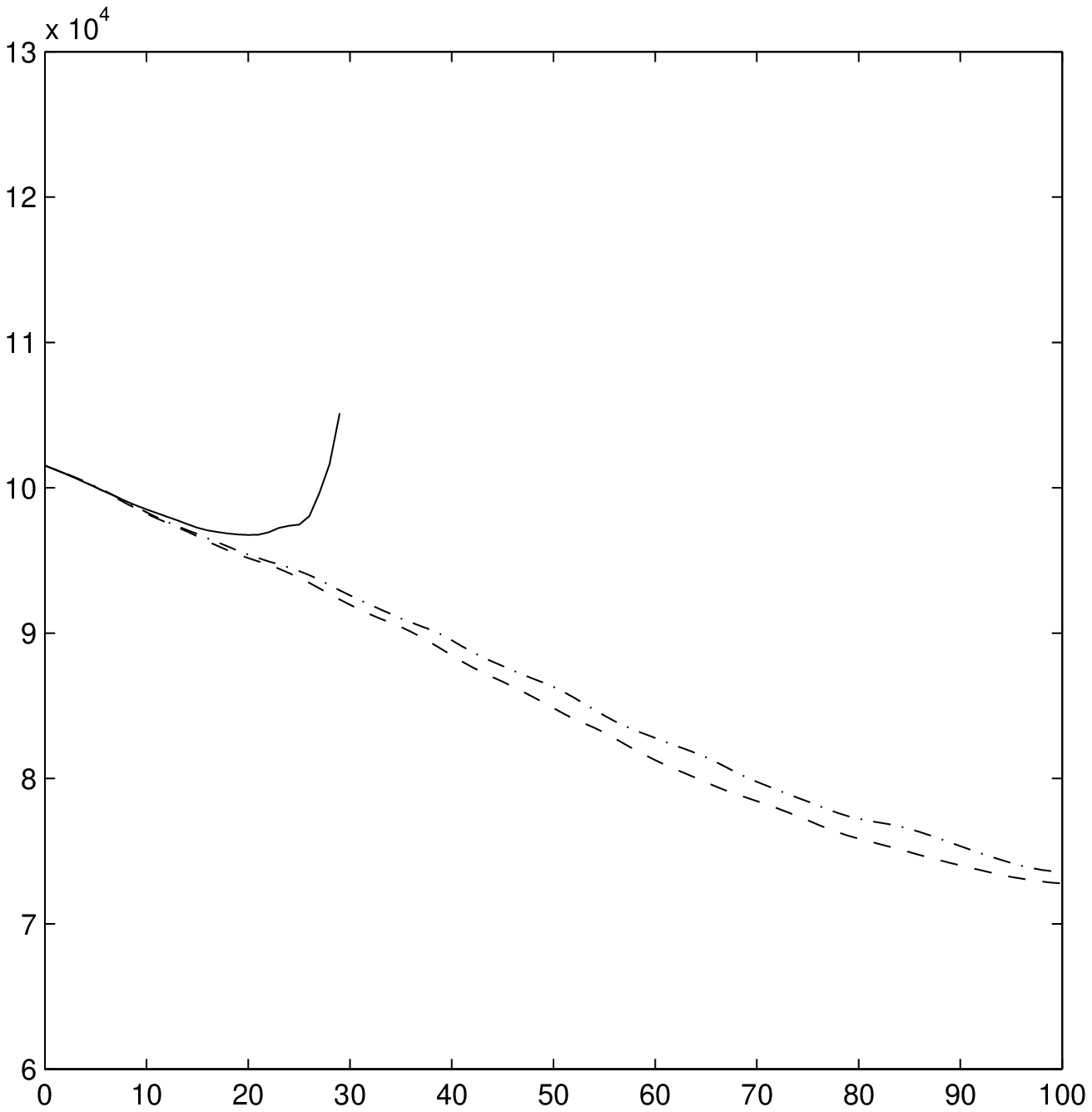,width=0.44\textwidth} (a)
\psfig{figure=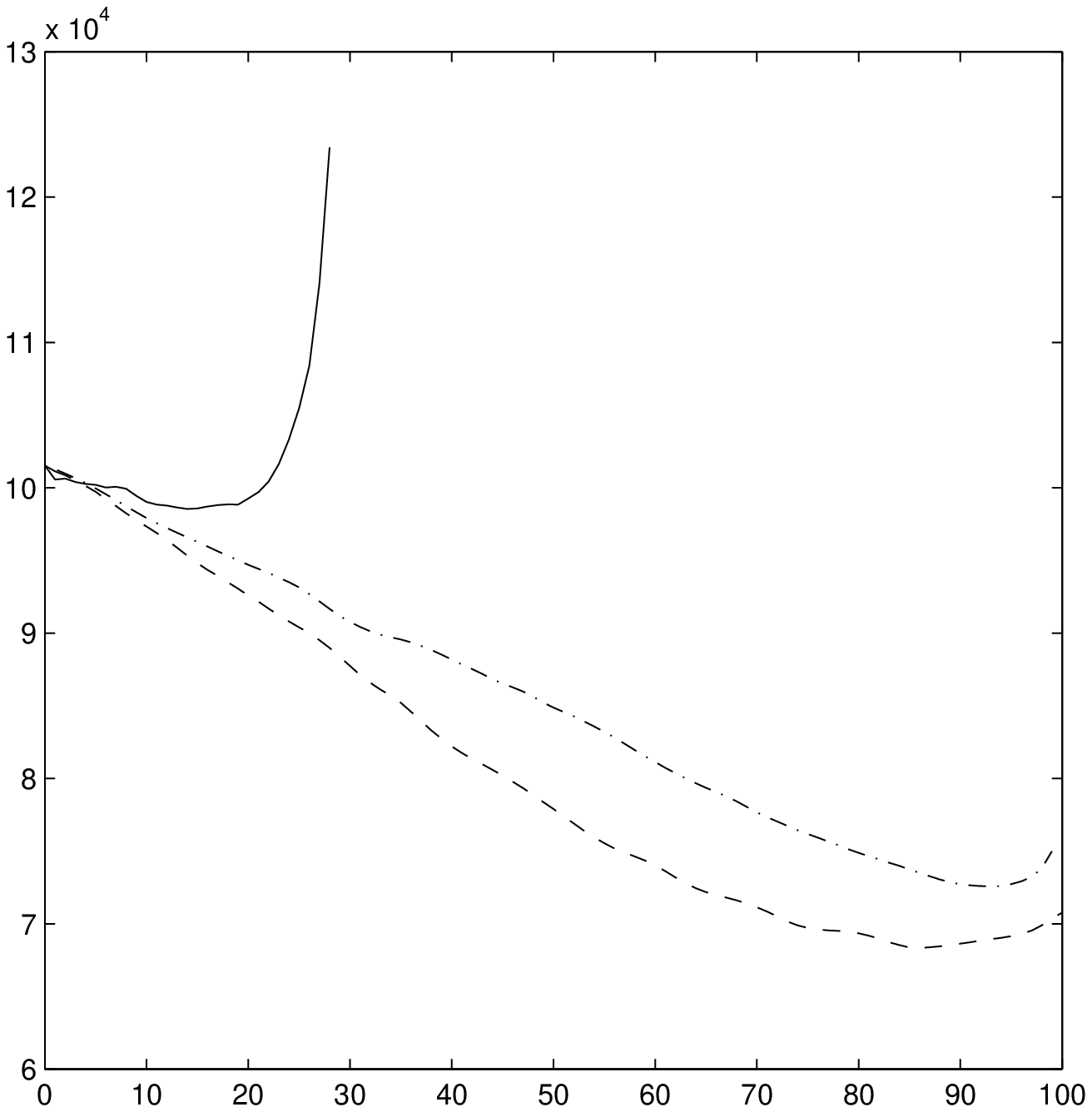,width=0.44\textwidth} (b)
}

\caption{Evolution of resolved kinetic energy for the nonlinear 
gradient model (solid)
and the filtered nonlinear gradient model, using $\kappa=1$ (dashed) 
and $\kappa=2$
(dash-dotted) (a). In (b) we show the corresponding results obtained 
with the unfiltered
(solid) and filtered full \lesa model. These instabilities are 
expected on grounds
discussed in subsection \ref{sec2-4}.}

\label{explfilt}

\end{figure}

The question arises
whether the explicit filtering can stabilize the simulations on this 
reference grid.
In figure~\ref{explfilt} we compiled predictions for the kinetic 
energy, obtained with
the nonlinear gradient and the full \lesa model, both without and 
with explicit filtering,
at different values of the ratio $\kappa$. We notice that the 
explicit filtering is
essential in order to maintain stability of the simulation. It 
appears that the unfiltered
\lesa model is even slightly more unstable than the unfiltered 
nonlinear gradient model.
We also considered these models at a higher resolution of $64^{3}$ 
grid-points. Consistent
with the analysis in section~\ref{sec2-4} the instability becomes 
stronger if the grid is
refined while keeping the LES filter-width $\Delta$ constant. It is 
seen that the value of
$\kappa$, which defines the width of the explicit filter relative to 
the width of the LES
filter, has a comparably small effect on the predictions of the 
nonlinear gradient model.
The instabilities which arise when using the full \lesa model, appear 
somewhat stronger
and,  e.g., $E$ even increases in the turbulent regime, despite the 
explicit filtering.
This indicates a marginally unstable simulation, and the situation 
improves when $\kappa$
is increased.

\subsubsection{Reference \lesa predictions}

Some basic predictions obtained using the three \lesa models will be 
presented next.
These predictions will contain errors because of shortcomings in the subgrid
parameterizations and the numerical treatment. These aspects will be 
focused upon in the
next two subsections respectively; here it is our aim to provide an 
impression of the
predictions under numerical conditions that are fairly common in 
present-day LES, e.g.
$\Delta/h=2$.

\begin{figure}[htb]

\centering{
{\psfig{figure=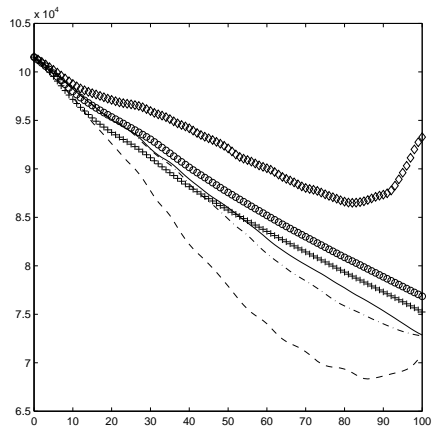,width=0.55\textwidth}}
}

\caption{Evolution of the resolved kinetic energy $E$ comparing the 
following models: \lesa (dashed), Leray (solid), filtered
nonlinear gradient model (dash-dotted), dynamic mixed (+), dynamic 
eddy-viscosity (o)
and no-model ($\diamond$). We used $\kappa=1$.}

\label{lesa_ekin}

\end{figure}

In figure~\ref{lesa_ekin} we compare the evolution of the resolved 
turbulent kinetic
energy $E$ for a number of subgrid models. We included not only 
predictions corresponding
to the three \lesa models, but also the dynamic mixed model, the 
dynamic eddy-viscosity
model and the simulation without any subgrid model at all. The 
subgrid models provide a
significant improvement compared to the case without a model. From 
previous simulations we
know that a fairly close agreement exists between filtered DNS data 
and the dynamic
models, as shown in figure~\ref{lesfig} (see \cite{vreman_jfm} for 
more details). Using
the dynamic predictions as point of reference here as well, we notice 
that the Leray and
the filtered nonlinear gradient model provide more accurate 
predictions than the full
\lesa model. We also considered the Bardina model and observed that 
the predictions are
virtually identical to those obtained with the filtered nonlinear 
gradient model. The
Smagorinsky model at $C_{S}=0.17$ was used as well and showed too 
strong dissipation.
\begin{figure}[htb]

\centering{
{\psfig{figure=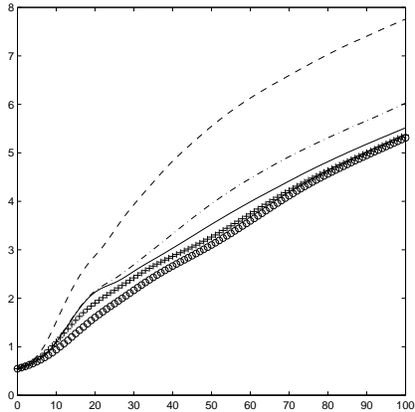,width=0.55\textwidth}}
}

\caption{Evolution of the resolved momentum thickness $\delta$ 
comparing the following
models: \lesa (dashed), Leray (solid), filtered nonlinear gradient 
model (dash-dotted),
dynamic mixed (+), dynamic eddy-viscosity (o). We used $\kappa=1$.}

\label{lesa_momthi}

\end{figure}

The momentum thickness $\delta$ is shown in figure~\ref{lesa_momthi}.
The prediction of $\delta$ from the full \lesa model is much higher 
than those obtained
with the other subgrid models and compared to the dynamic model 
predictions as point of
reference, it appears too high. The predictions of the Bardina 
similarity model again
coincide with the filtered nonlinear gradient model, and these 
predictions are somewhat
larger than arise from the Leray model. All the \lesa models predict 
$\delta$ larger than
the dynamic models. Since the dynamic predictions slightly 
underestimate $\delta$
according to \cite{vreman_jfm}, it appears that the Leray model and 
the filtered nonlinear
gradient model predict $\delta$ more accurately, compared to filtered 
DNS results, than
the other models.

\begin{figure}[htb]

\centering{
{\psfig{figure=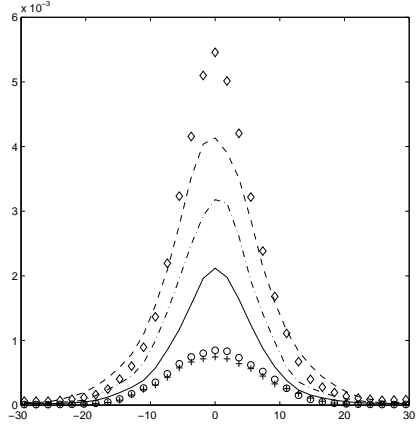,width=0.55\textwidth}}
}

\caption{Comparison of the Reynolds stress
$-\langle w_{1}w_{2} \rangle$ at $t=70$: \lesa (dashed), Leray 
(solid), filtered
nonlinear gradient model (dash-dotted), dynamic mixed (+), dynamic 
eddy-viscosity (o) and
no-model ($\diamond$). We used $\kappa=1$.}

\label{lesa_reynold}

\end{figure}

In figure~\ref{lesa_reynold} we collected the
Reynolds stress $-\langle w_{1}w_{2} \rangle$. We observe that all three
\lesa models predict a considerably higher level of fluctuations 
compared to the dynamic
models. The full \lesa model predicts levels of fluctuation close to those
obtained from the simulation without any subgrid model, suggesting 
that this model
introduces too many small scales into the solution. Likewise, the 
filtered gradient model
generates high levels of fluctuations, while the Leray model is much 
closer to the levels
of fluctuation that are found using the dynamic models.

\begin{figure}[htb]

\centering{
{\psfig{figure=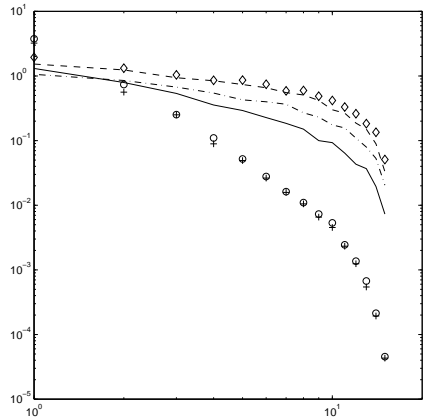,width=0.55\textwidth}}
}

\caption{Comparison of the streamwise energy spectrum $A(k)$ at 
$t=80$: \lesa (dashed),
Leray (solid), filtered nonlinear gradient model (dash-dotted), 
dynamic mixed (+), dynamic
eddy-viscosity (o) and no-model ($\diamond$). We used $\kappa=1$.}

\label{lesa_spect}

\end{figure}

We consider the streamwise kinetic energy spectrum in the turbulent 
regime  at $t=80$, in
figure~\ref{lesa_spect}. We observe a clear separation of the 
predictions in two groups.
The two dynamic models show a strong reduction of the smaller scales. 
In contrast, the
full \lesa model displays a spectrum that is quite close to the 
spectrum of the simulation
without any subgrid model. This situation improves significantly for 
the filtered
nonlinear gradient model and finally, the Leray model provides the 
largest attenuation of
the small scales among the three \lesa models.

In summary, the simulations suggest that the full \lesa model does 
not sufficiently
reduce the resolved kinetic energy, leads to too large 
momentum-thickness and too high
levels of fluctuation, which is apparent in the spectrum at small 
scales and snapshots of
the solution. The filtered nonlinear gradient model performs better 
than the full \lesa
model but also over-predicts the smaller scales. In contrast to these 
two models, the
Leray model, appears to predict the energy decay properly, shows accurate
momentum-thicknesses and apparently reliable levels of turbulence 
intensities, as shown
also in the spectrum and in snapshots of the solution. In order to 
better understand these
predictions we turn to the resolved kinetic energy dynamics in the 
next subsection and
consider the contribution of the individual terms in the models.

\subsection{Resolved kinetic energy dynamics}
\label{energy}

In this section we consider the evolution of the resolved kinetic energy and
determine the type and magnitude of the various subgrid 
contributions. The evolution of
$E$ is governed by
\bea
\pr_{t} E
&=&
\int_{\Omega} \{ \frac{1}{Re}\ub_{i} \pr_{j} {\overline{\sigma_{ij}}}
  - \ub_{i} \pr_{j} \tau_{ij} \} ~ d{\bf{x}} \nonumber \\
&=&
\int_{\Omega} \{-\frac{1}{2Re} \ovl{\sigma_{ij}}~ \ovl{\sigma_{ij}}
+ \tau_{ij} \pr_{j} \ub_{i}\} ~ d {\bf{x}}
\ena
where use was made of the identity
$\ovl{\sigma_{ij}}\pr_{j}\ub_{i}=\frac12 
{\ovl{\sigma_{ij}}}~{\ovl{\sigma_{ij}}}$.
The predicted kinetic energy evolution, corresponding to a given LES 
model, emerges by
replacing the turbulent stress tensor by its subgrid scale model. We 
notice that the
dynamics of $E$ is governed by a purely dissipative term arising from 
the molecular
dissipation and a term that is associated with the subgrid model. We 
will consider the
resolved energy dynamics both for the coarse reference grid of 
$32^{3}$ grid points and a
much finer simulation in which we use $96^{3}$. The latter 
simulations use the same
filter-width $\Delta=L/16$ but correspond to a much higher subgrid resolution
$\Delta=6h_{LES}$. In this way we can clarify some of the dynamics 
observed on the coarse
grid as well as obtain an impression of the actual dynamical 
consequences associated with
the subgrid model.

\begin{figure}[htb]

\centering{
{\psfig{figure=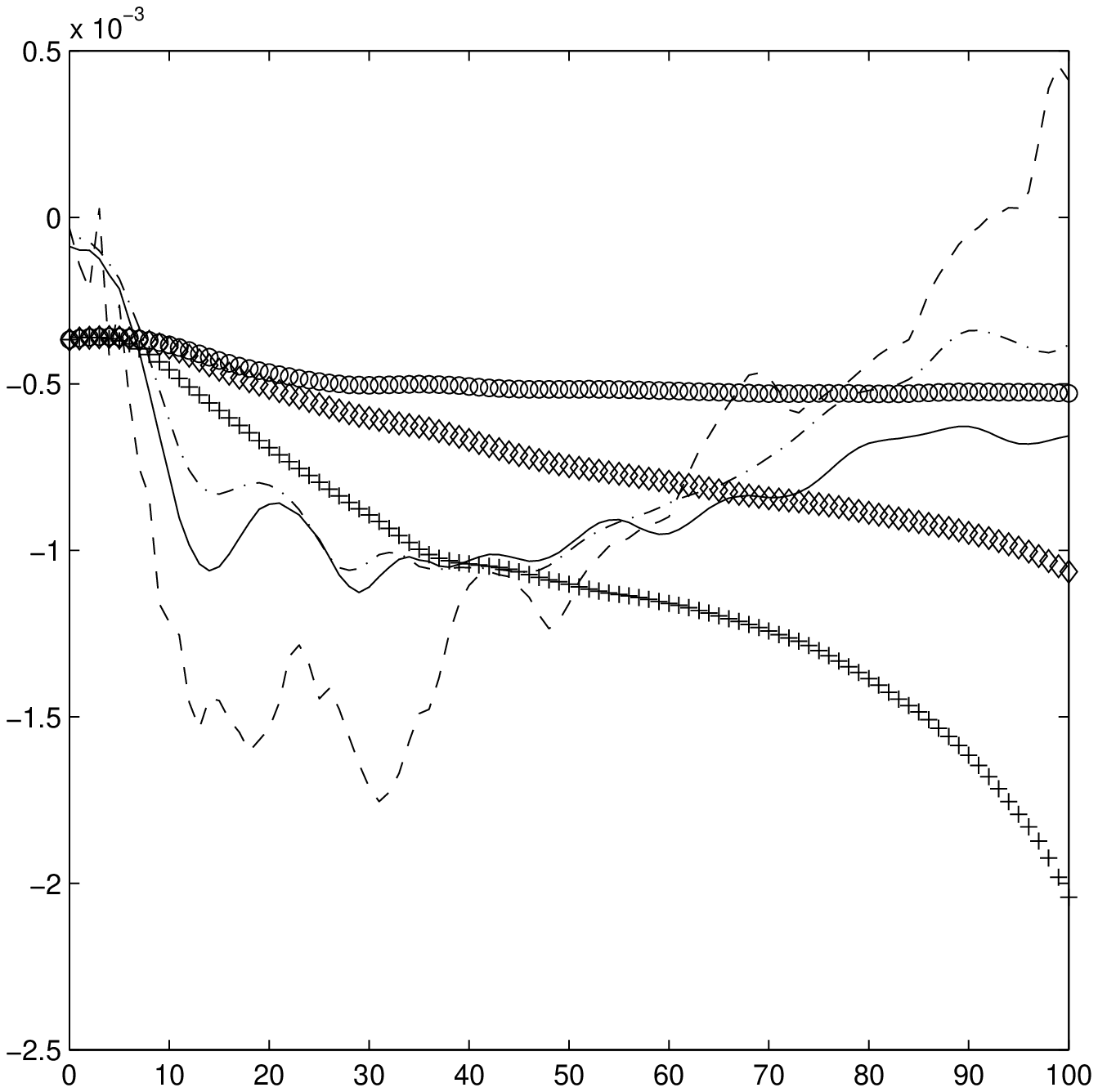,width=0.44\textwidth}} (a)
{\psfig{figure=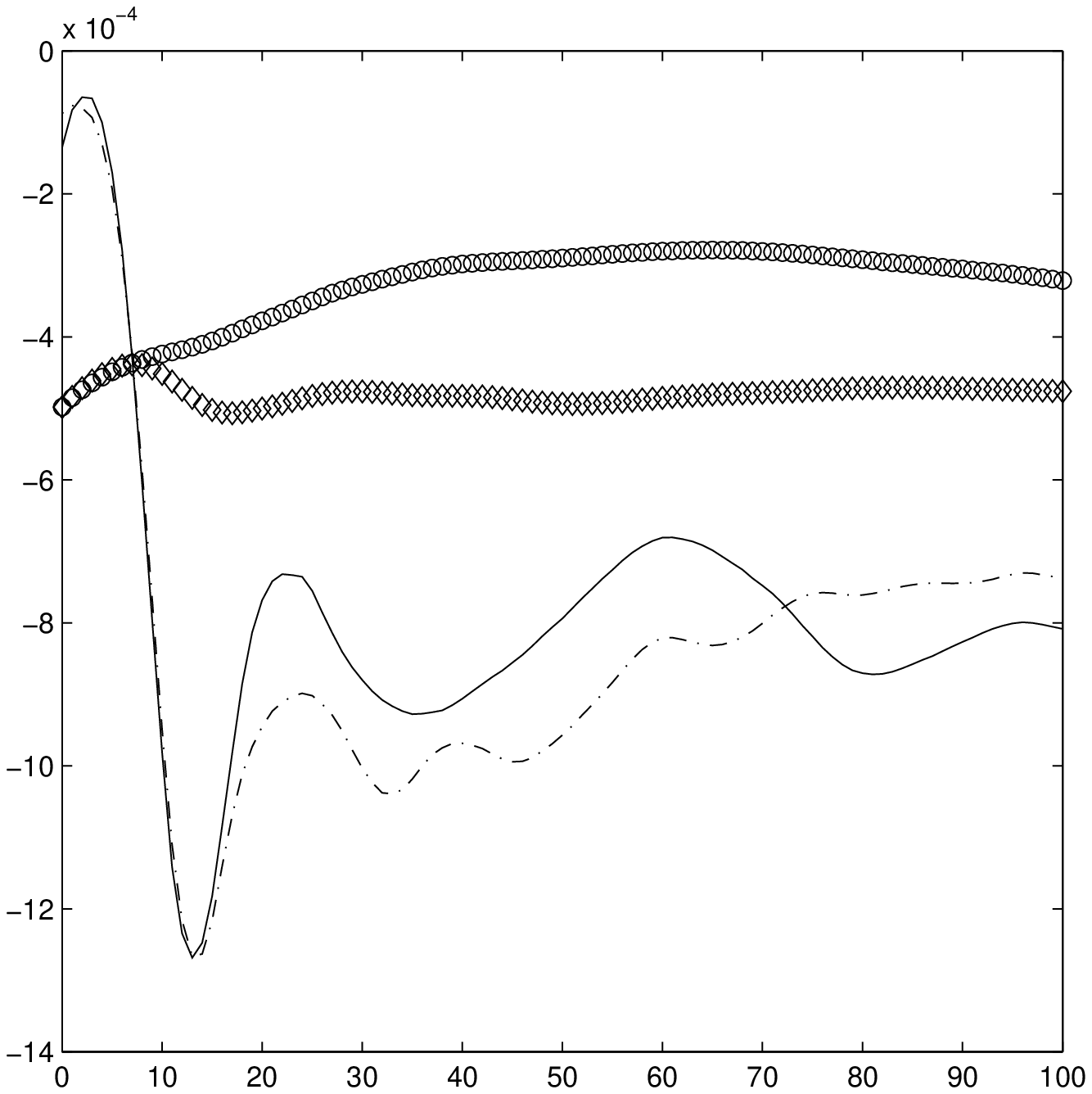,width=0.44\textwidth}} (b)
}

\caption{Resolved kinetic energy rate contributions as a function of time:
\lesa (dashed line $\pr_{t}E_{m}$, +: $\pr_{t}E_{v}$), Leray (solid 
line: $\pr_{t}E_{m}$,
o: $\pr_{t}E_{v}$ ), filtered nonlinear gradient model (dash-dotted 
line $\pr_{t}E_{m}$,
$\diamond$: $\pr_{t}E_{v}$). We used $\kappa=1$ and show the results 
for the $32^{3}$ grid
in (a) and for the $96^{3}$ grid in (b).}

\label{lesa_ekin_total}

\end{figure}

In figure~\ref{lesa_ekin_total} we show the total viscous and subgrid 
contributions to
$\pr_{t}E$, denoted $\pr_{t}E_v$ and $\pr_{t}E_m$, respectively. We 
notice that on the
$32^{3}$ grid the viscous contribution corresponding to the Leray 
model is quite constant
in the turbulent regime and the subgrid contribution gradually 
becomes of the same order of
magnitude. For the filtered nonlinear gradient model we observe a 
proper dissipation of
energy, but slightly less than the Leray model. The corresponding viscous flux
contribution increases considerably in the turbulent regime. Finally, 
for the full \lesa
model we observe that the subgrid contribution not only becomes less 
important in the
turbulent regime but even changes sign. This can readily be associated with the
overestimated small scale contributions in the solution, as shown in the 
previous
subsection. For the better resolved LES the results of the Leray 
model and the filtered
nonlinear gradient model are quite comparable and appear more 
predictable. Moreover, all
subgrid fluxes are seen to settle and oscillate around some nonzero 
values, indicating
perhaps a more regular self-similar development of the mixing layer 
in the turbulent
regime. The full \lesa model was found to become unstable around
$t=70$, at this high subgrid resolution. Apparently, the explicit 
filtering, which was
found to be essential in the previous section, in order to stabilize 
the simulation on the
coarse grid, is not damping sufficiently well to maintain stability 
of the \lesa model at
increased subgrid resolution.

\begin{figure}[htb]

\centering{
{\psfig{figure=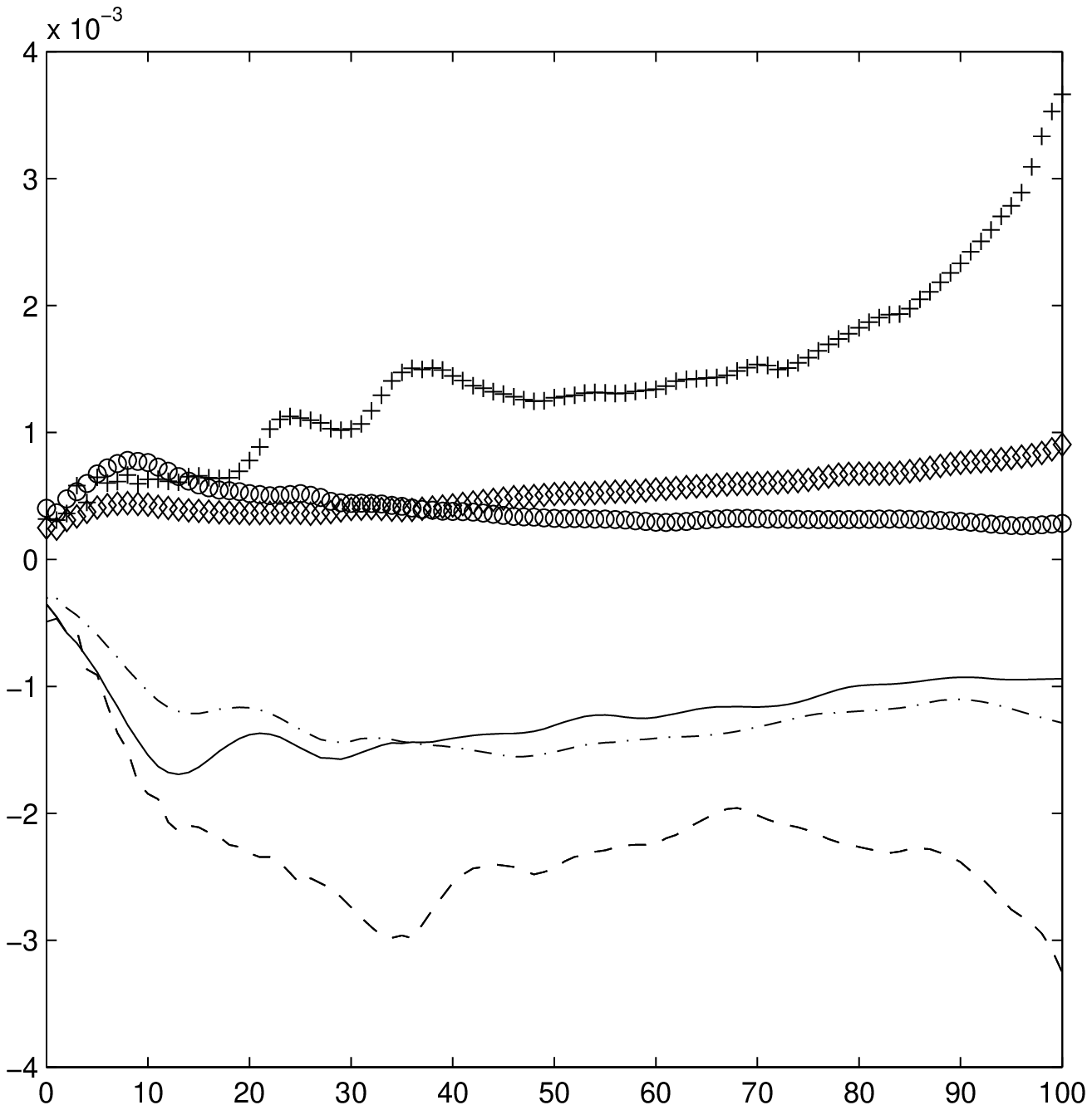,width=0.44\textwidth}} (a)
{\psfig{figure=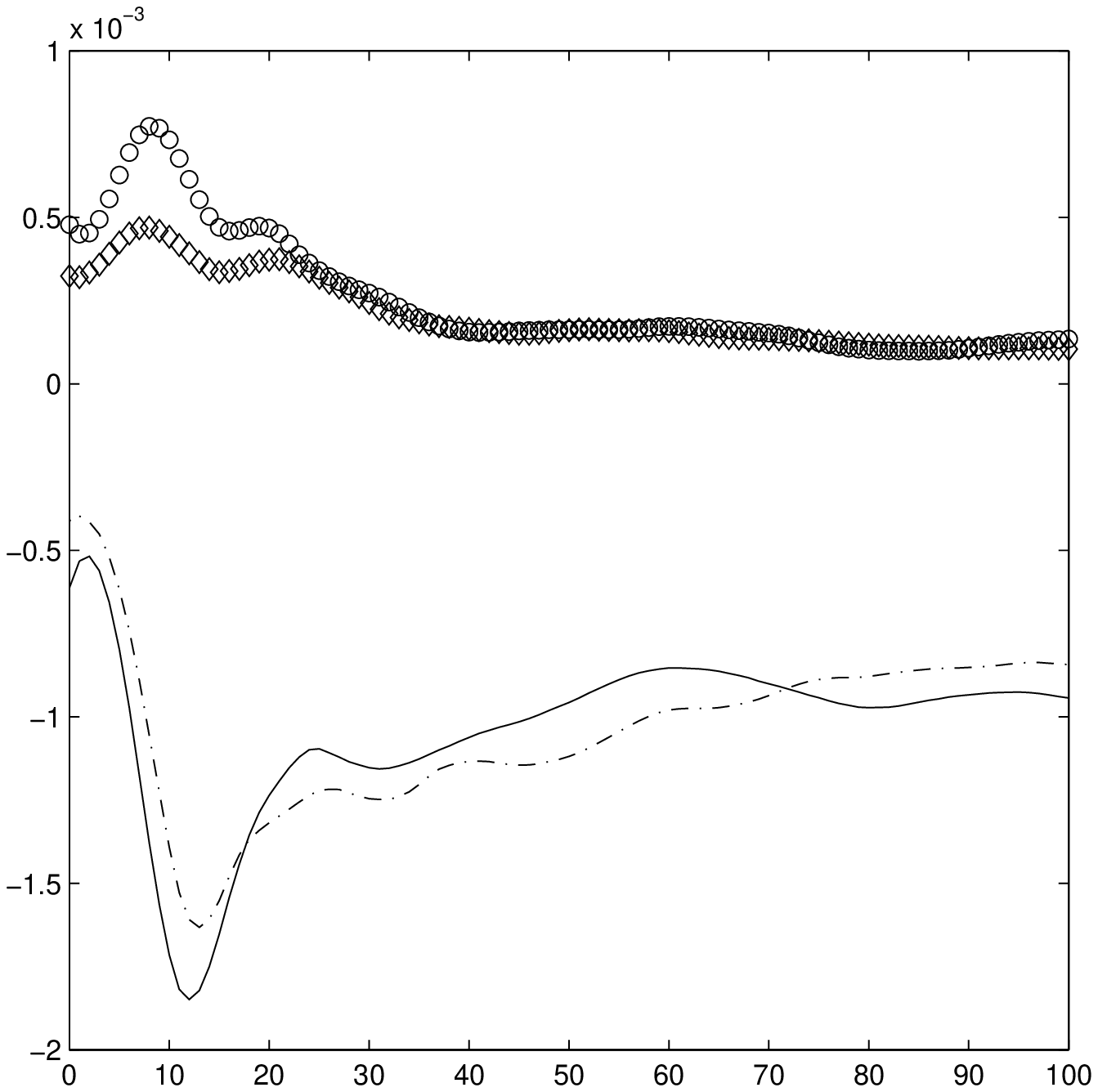,width=0.44\textwidth}} (b)
}

\caption{Resolved kinetic energy rate contributions: \lesa (dashed 
line $P_{f}$, +:
$P_{b}$), Leray (solid line: $P_{f}$, o: $P_{b}$ ), filtered 
nonlinear gradient model
(dash-dotted line $P_{f}$, $\diamond$: $P_{b}$). We used $\kappa=1$ 
and show the results
for the
$32^{3}$ grid (a) and the $96^{3}$ grid (b).}

\label{lesa_ekin_forw_back}

\end{figure}

To further analyse the dynamical behavior, we can look at splitting the
subgrid contribution into a positive, i.e., forward scatter or 
dissipative, contribution
and a negative, i.e., backward scatter or reactive contribution. To 
formalize this
splitting, we introduce
\be
P_{b}(f)=\int_{\Omega}\frac12(f + |f|)~d{\bf{x}}
~,~~
P_{f}(f)= \int_{\Omega}\frac12(f - |f|)~d{\bf{x}}
\en
to measure the amount of back-scatter ($P_{b}$) and forward scatter 
of energy ($P_{f}$)
associated with a term represented by $f$. In 
figure~\ref{lesa_ekin_forw_back} we
collected the forward and backward scatter contributions for the 
\lesa models. We observe
that all these models predict both forward and backward scatter of 
energy, which sets them
apart from simple eddy-viscosity models that only provide forward 
scatter. On the coarse
grid ($32^{3}$) the Leray model and the filtered nonlinear gradient 
model compare fairly
well. The full \lesa model, however, shows a large amount of 
back-scatter in the turbulent
regime and a likewise increased importance of forward scatter. On the 
finer grid the Leray
and filtered nonlinear gradient model show a balance between forward 
and backward scatter
in the turbulent regime.

\begin{figure}[htb]

\centering{
{\psfig{figure=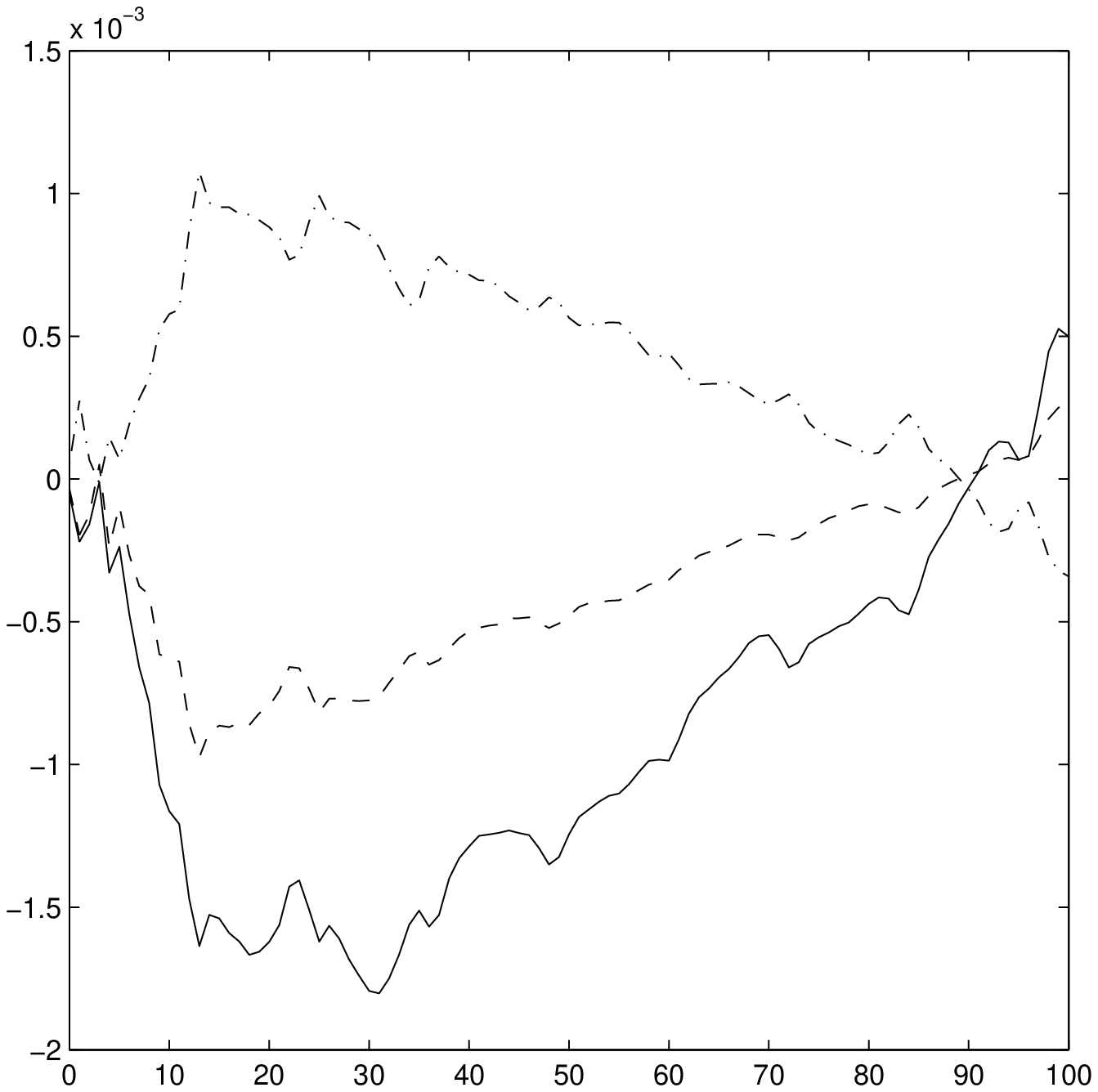,width=0.44\textwidth}} (a)
{\psfig{figure=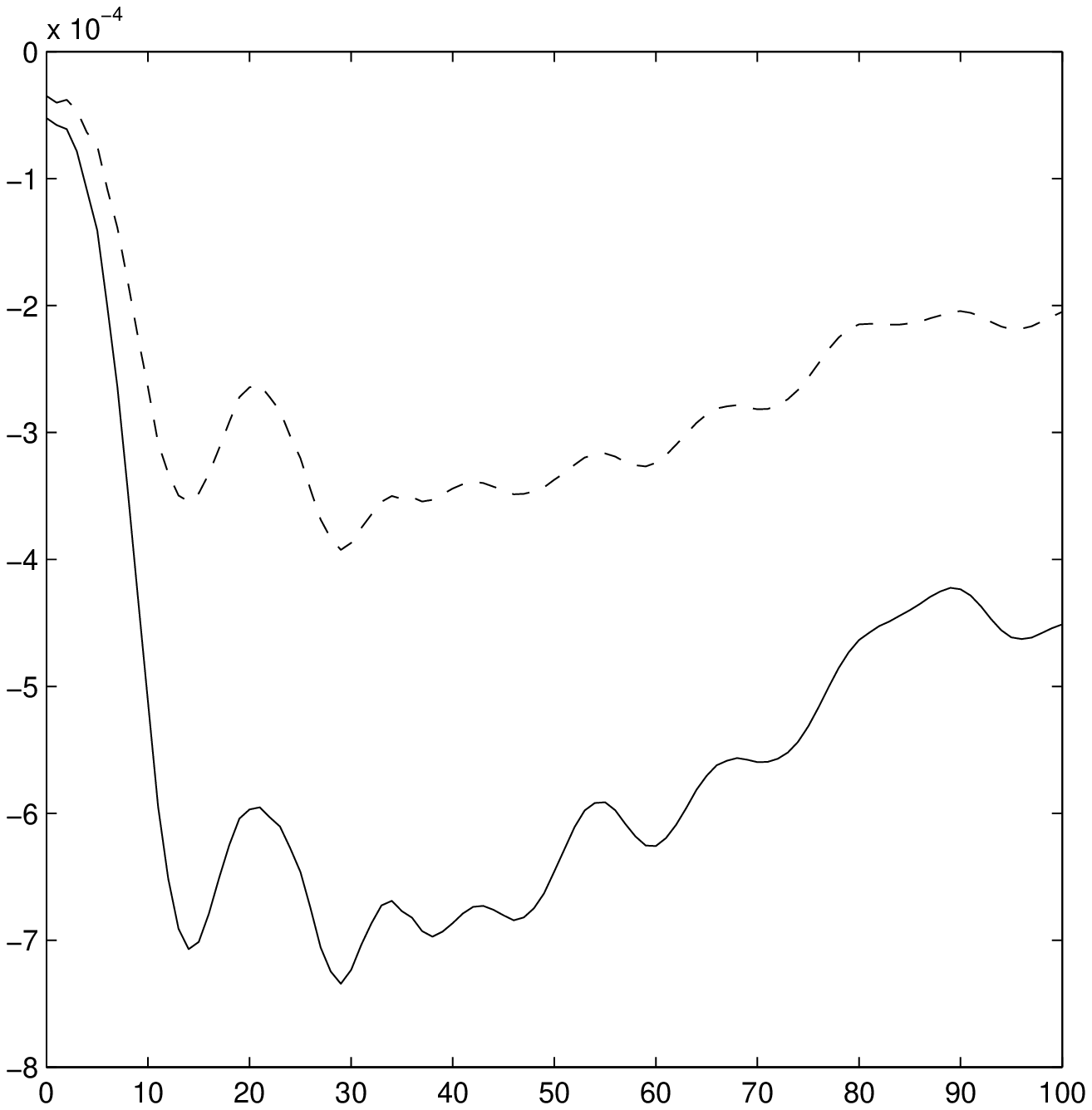,width=0.44\textwidth}} (b)
}

\caption{Resolved kinetic energy rate contributions: full \lesa shown 
in (a) (solid
${\ovl{A}}$-term, dashed ${\ovl{B}}$-term, dash-dotted 
${\ovl{C}}$-term), Leray model
shown in (b):(solid ${\ovl{A}}$-term, dashed ${\ovl{B}}$-term) We 
used $\kappa=1$ and show
the results for the $32^{3}$ grid.}

\label{lesa_ekin_decomp}

\end{figure}

A third decomposition of the total contribution arises in terms of 
the individual
subgrid-terms. If we consider,  e.g., the full \lesa model, written as
$m_{ij}^{\alpha}={\ovl{A_{ij}}}+{\ovl{B_{ij}}}-{\ovl{C_{ij}}}$ we may write
\be
\pr_{t}E=\pr_{t}E_{v}+\pr_{t}E_{A}+\pr_{t}E_{B}-\pr_{t}E_{C}
\en
with individual contributions due to the viscous fluxes, and the $A-B-C$ terms
respectively;
\be
\pr_{t} E_{v}
=
\int_{\Omega} -\frac{1}{2Re}
\ovl{\sigma_{ij}}~ \ovl{\sigma_{ij}}
~ d {\bf{x}}~,~~
\pr_{t} E_{A}
=
  \int_{\Omega} {\ovl{A_{ij}}} \pr_{j}\ub_{i} ~ d {\bf{x}}
\en
and similarly for the other terms. In figure~\ref{lesa_ekin_decomp} 
we collected the
detailed energy-dynamics decomposition corresponding to the two terms 
which make up the
Leray model and the three terms that constitute the full \lesa model. 
Notice that
figure~\ref{lesa_ekin_total} already contains the single contribution 
of the filtered
nonlinear gradient model. The Leray model is seen to be composed of 
two terms that both
dissipate energy. The full \lesa model behaves less regular and we 
observe that the
dissipative ${\ovl{B}}$ contribution is nearly canceled by the reactive
${\ovl{C}}$ contribution.

 From this analysis of the resolved energy dynamics it seems that the 
Leray model and the
filtered nonlinear gradient model provide more accurate results and 
the internal
functioning of these models increases the robustness of the model. We 
also applied the
Leray model to a flow at a ten times higher Reynolds number. Although 
these latter
results are still preliminary, it seems that the Leray model provides 
reliable results,
even in such very turbulent flows. Further analysis of this regime is 
needed though and
this will be published elsewhere \cite{geurts_holm}.

In the next section we use the well resolved LES predictions in 
combination with the
coarse grid simulation results to assess the influence of the spatial 
discretization
scheme on the evolution of the flow.

\subsection{Toward grid-independent \lesa}
\label{gridindependent}

The reference simulations considered above, are executed on a fairly 
coarse grid which
corresponds to a ratio between filter-width $\Delta$ and grid-spacing 
$h$ of $\Delta/h=2$.
In many present-day LES, even a ratio of $\Delta/h=1$ is frequently 
used. These choices
usually arise from considerations of available computational 
resources, but at the same
time imply that the smallest resolved scales of size on the order of 
$\Delta$ are not
accurately represented in the numerical treatment. Hence, there is a 
strong possibility
that the marginal subgrid resolution influences the dynamical 
properties of the simulated
flow.

In order to assess this discretization effect, we will compare the 
reference LES with
simulations at a much higher subgrid resolution. In this way, the 
effect of subgrid
modeling is better represented numerically, while it remains of the 
same magnitude as in
the coarse grid simulation. This allows to isolate the dynamic 
effects of the spatial
discretization in the modeled equations.

\begin{figure}[htb]

\centering{
{\psfig{figure=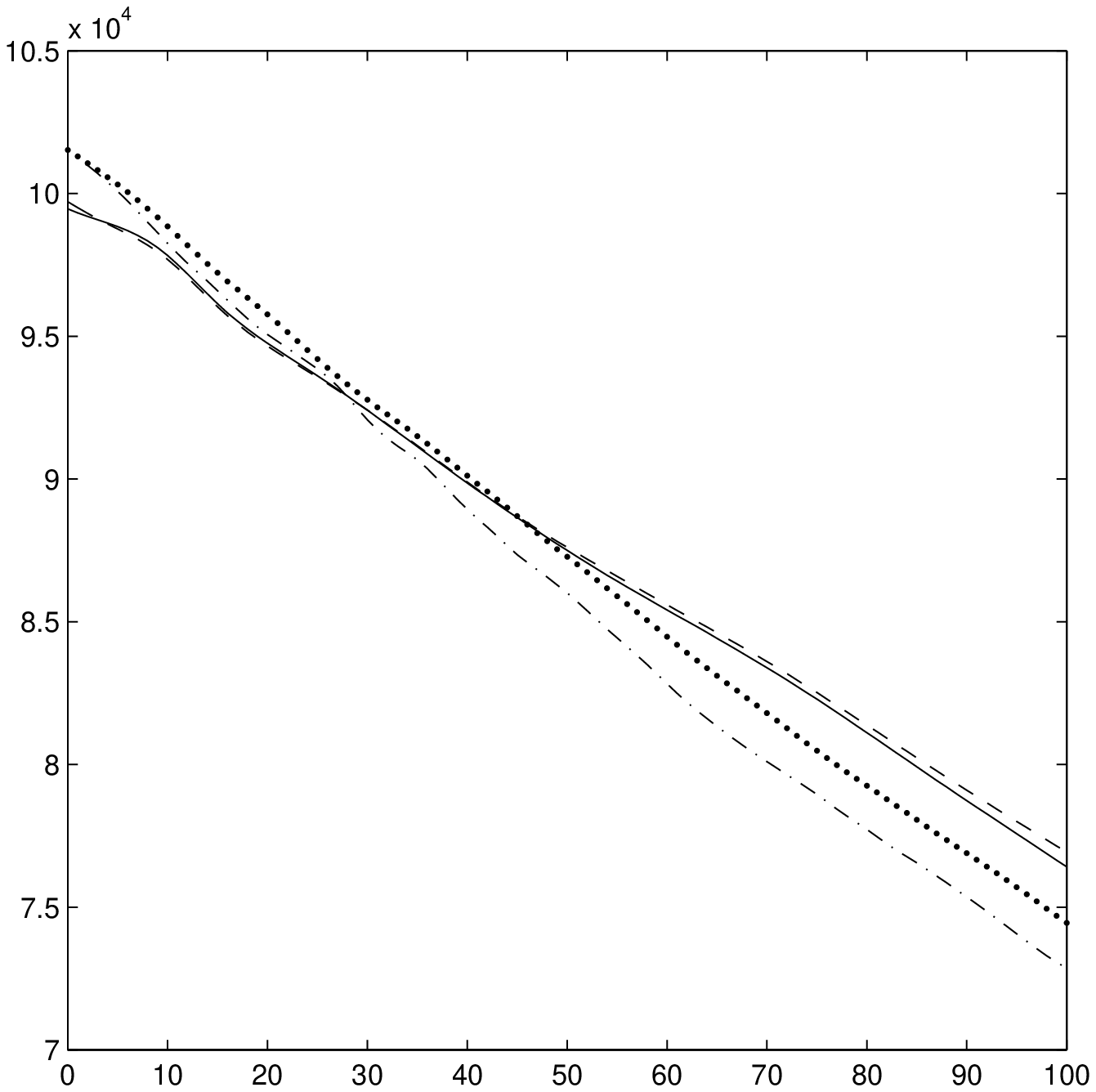,width=0.44\textwidth}} (a)
{\psfig{figure=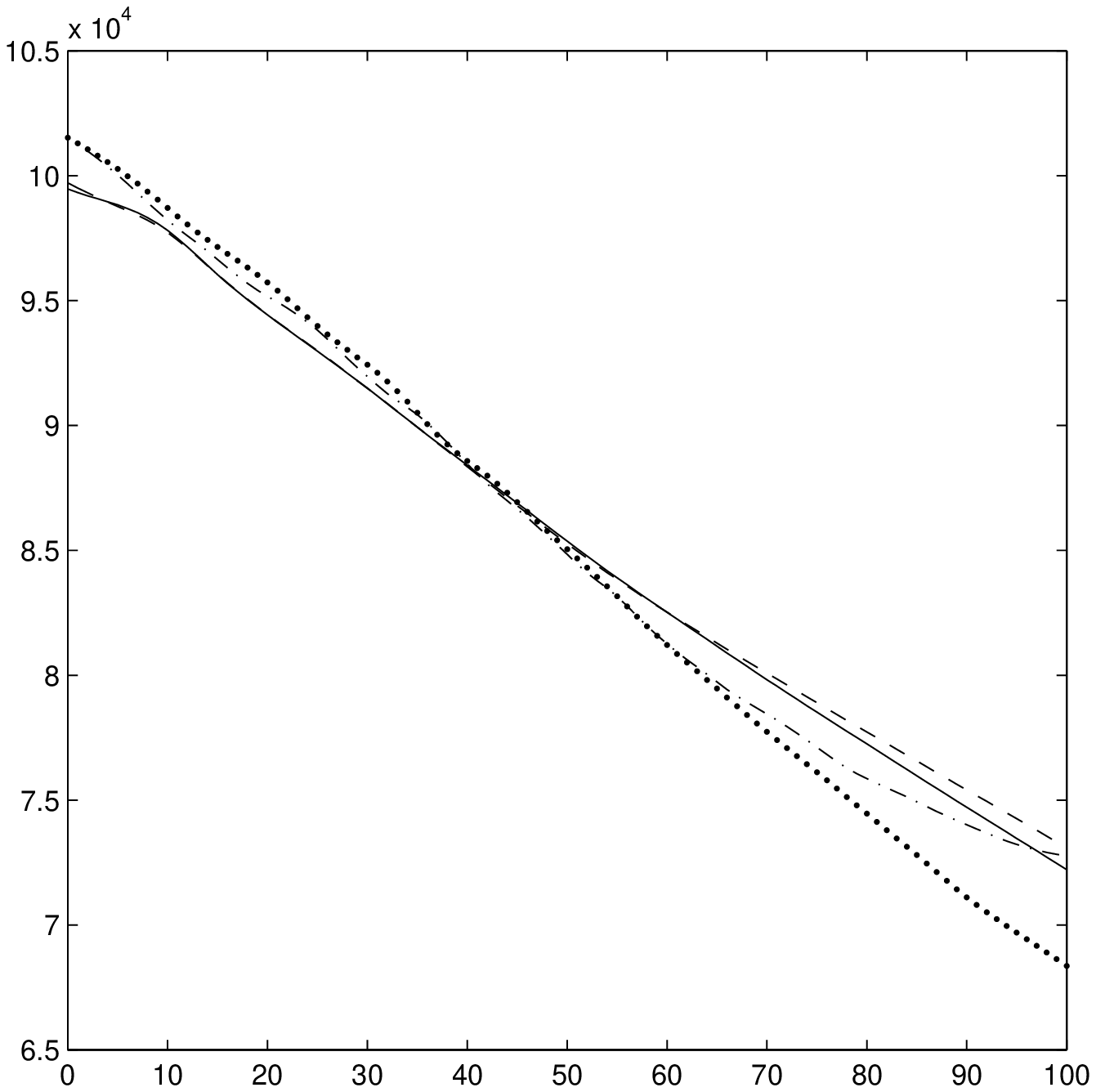,width=0.44\textwidth}} (b)
}

\caption{Resolved kinetic energy using the Leray model (a) and the 
filtered nonlinear
gradient model (b): solid line (4th order method, resolution 
$96^{3}$), dashed line (4th
order method, resolution $64^{3}$), dash-dotted line (4th order 
method, resolution
$32^{3}$) and dotted-line (2nd order method, resolution
$32^{3}$).}

\label{lesa_ekin_grids}

\end{figure}

We compare simulations on $32^{3}$, $64^{3}$ and $96^{3}$ grid-points 
and focus our
attention on the Leray model and the filtered nonlinear gradient model. In
figure~\ref{lesa_ekin_grids} we compare the predicted resolved 
kinetic energy obtained
with the second and the fourth order accurate spatial discretization 
method. The subgrid
resolution corresponding to these three grids is $\Delta/h=2$, 4 and 
6 respectively. We
observe a very close agreement between the predictions using the 
fourth order accurate
method and $\Delta/h=4$ and 6. This suggests that a mean flow 
quantity such as the
resolved kinetic energy is well represented using $\Delta/h=4$. 
Moreover, we notice that
on the coarsest grid, the accuracy of the prediction based on the 
second order method
compares closely to that obtained with the fourth order method. 
Apparently, if the dynamic
effects of the spatial discretization errors are quite large, a lower 
order method can be
competitive with a higher order method. For both models the 
reliability of the predictions
on the coarse grid are affected considerably by the coarseness of the 
subgrid resolution.
In both situations, and for both spatial discretizations the effect 
of the discretization
error is seen to enhance the reduction of the resolved kinetic energy.

\begin{figure}[htb]

\centering{
{\psfig{figure=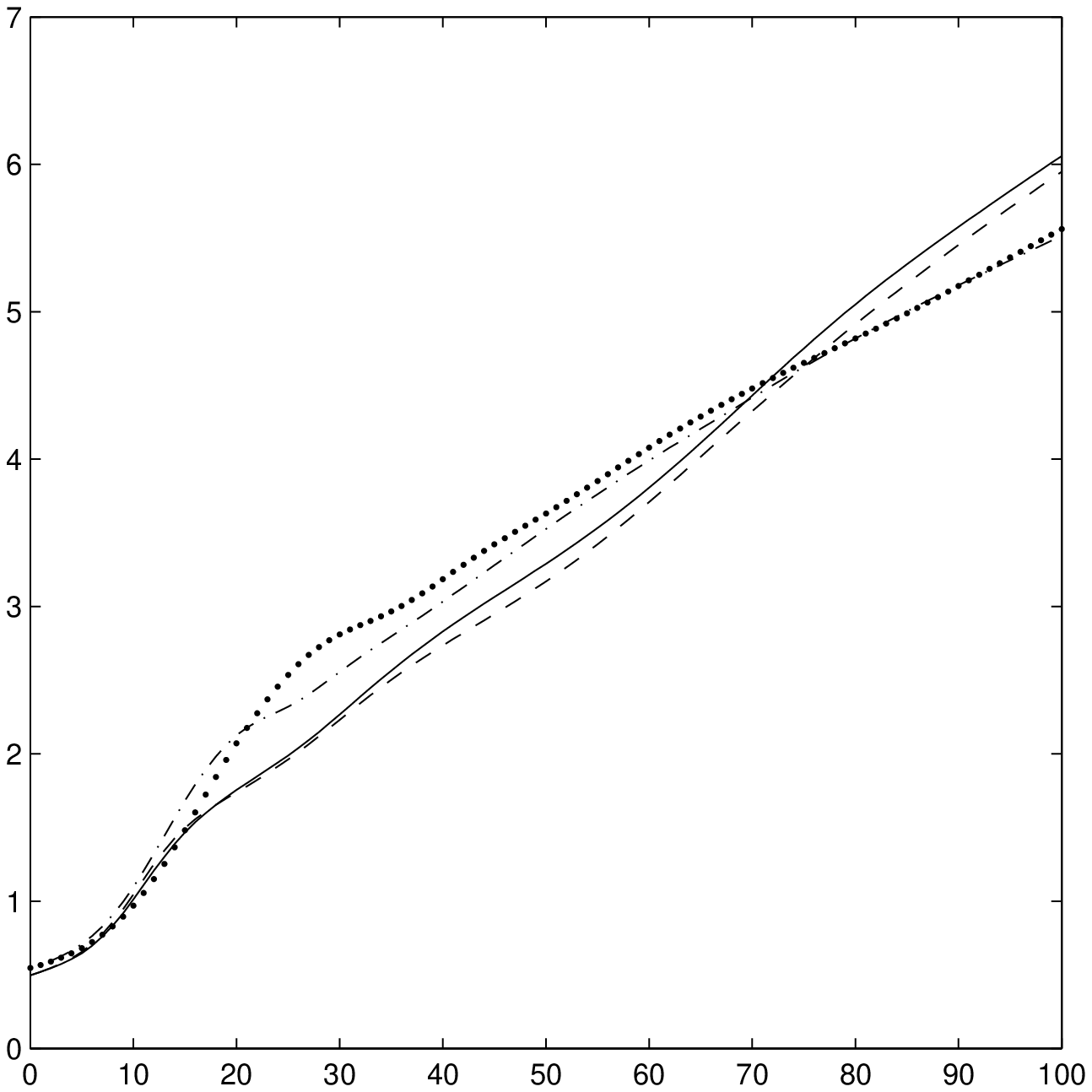,width=0.44\textwidth}} (a)
{\psfig{figure=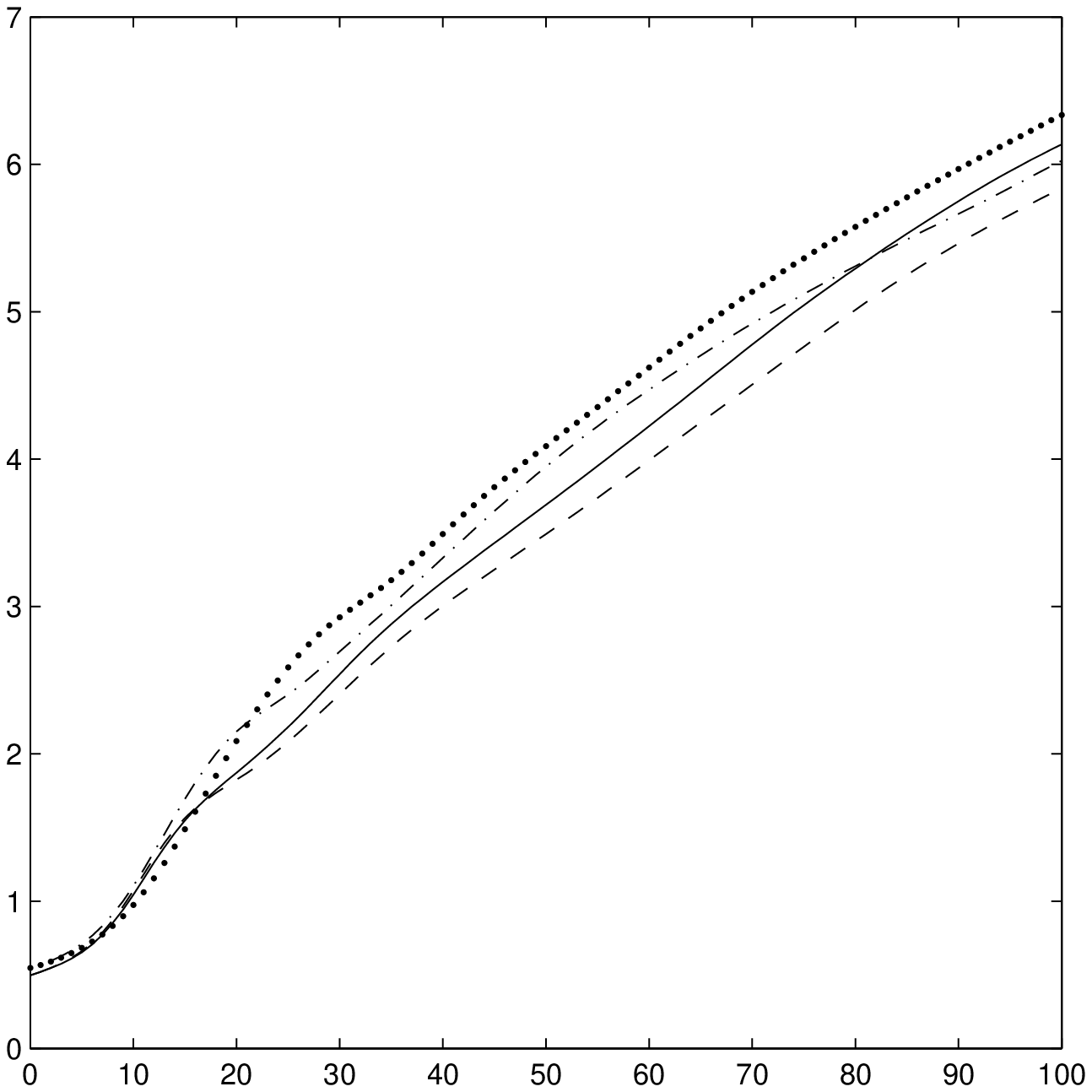,width=0.44\textwidth}} (b)
}

\caption{Resolved momentum-thickness using the Leray model (a) and 
the filtered nonlinear gradient model (b):
solid line (4th order method, resolution $96^{3}$), dashed line (4th 
order method,
resolution $64^{3}$), dash-dotted line (4th order method, resolution 
$32^{3}$) and dotted-line (2nd order method, resolution
$32^{3}$).}

\label{lesa_momthi_grids}

\end{figure}

To further establish the convergence, we show the momentum-thickness in
figure~\ref{lesa_momthi_grids}. We observe that the convergence is 
clear for the
Leray model and that a value of $\Delta/h=4$ corresponds to reliable 
predictions for both
subgrid models considered, although the sensitivity of the momentum 
thickness is larger
than that of the resolved kinetic energy. Regarding the results for 
the best resolved
simulations, we observe that the momentum-thickness develops very 
nearly linearly with
time in the Leray model, while a slight reduction of the growth-rate 
predicted by the
nonlinear gradient model is seen in the turbulent regime.

\begin{figure}[htb]

\centering{
{\psfig{figure=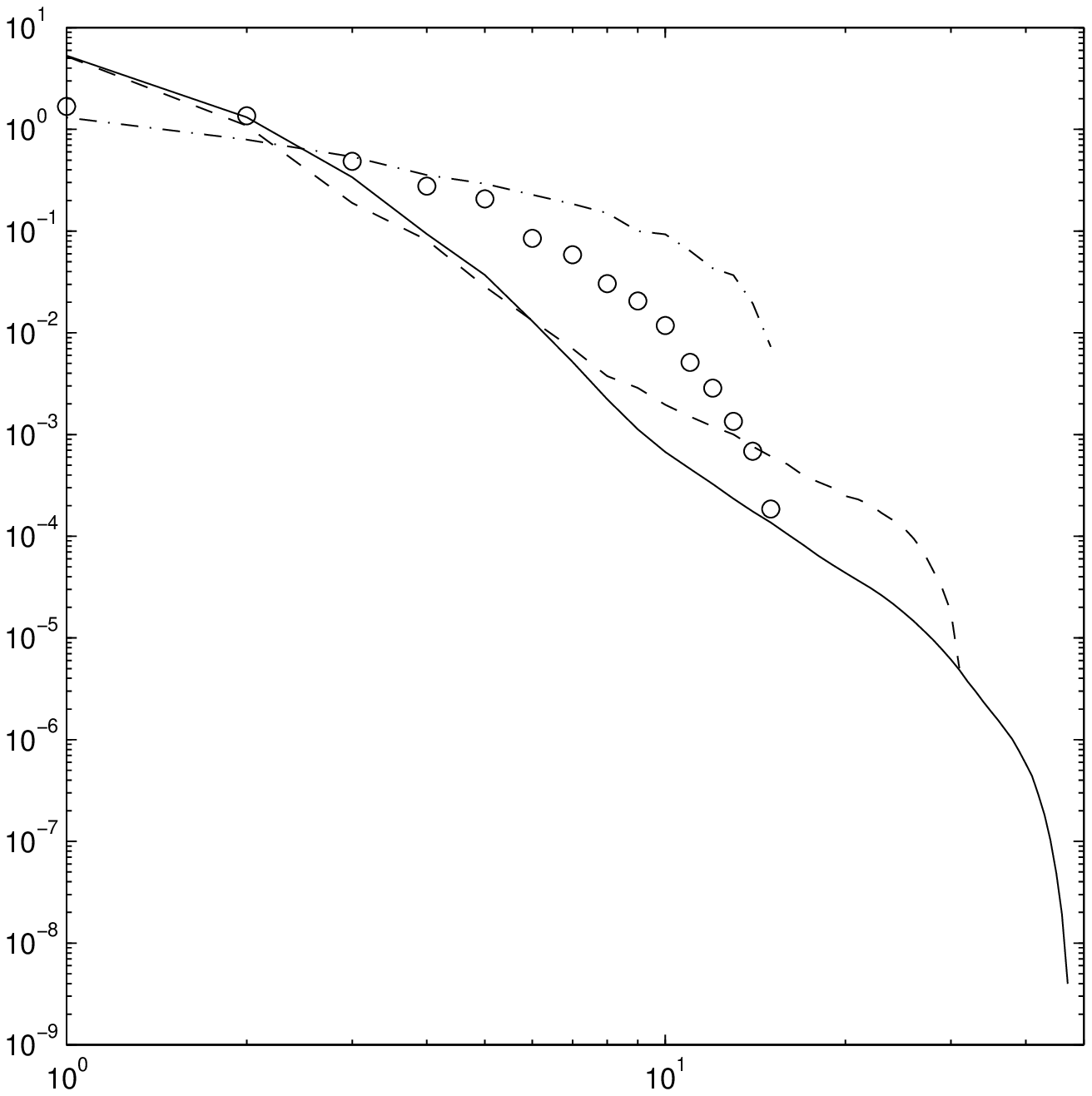,width=0.44\textwidth}} (a)
{\psfig{figure=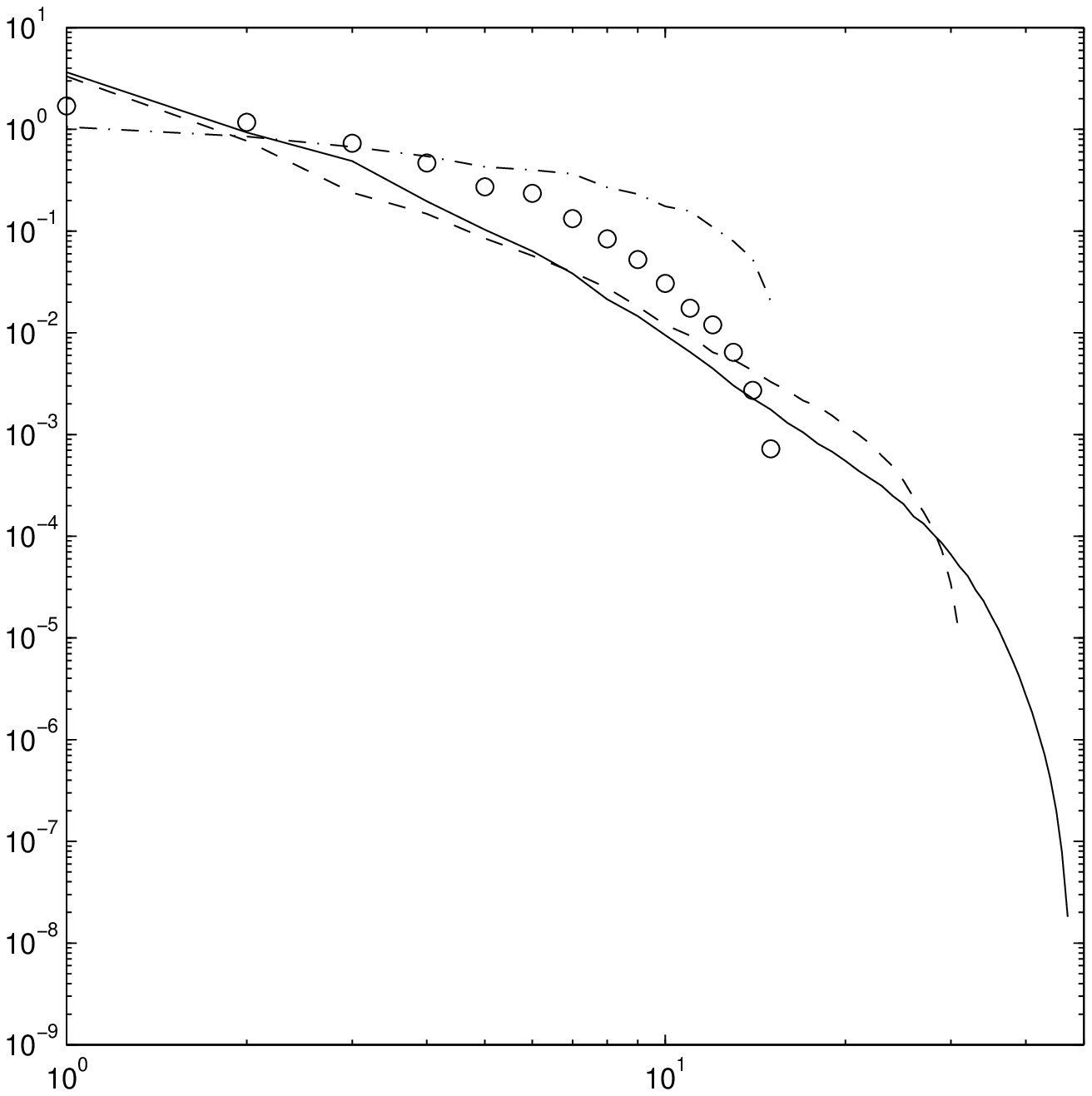,width=0.44\textwidth}} (b)
}

\caption{Streamwise kinetic energy spectrum using the Leray model (a) 
and the filtered
nonlinear gradient model (b): solid line (4th order method, 
resolution $96^{3}$), dashed
line (4th order method, resolution $64^{3}$), dash-dotted line (4th 
order method,
resolution $32^{3}$) and marker `o' (2nd order method, resolution
$32^{3}$).}

\label{lesa_spec_grids}

\end{figure}

Finally, we show the spectra obtained with the Leray and the filtered 
nonlinear gradient
model on the selected grids in figure~\ref{lesa_spec_grids}. We 
notice a general
resemblance between the results obtained with both models. As the 
subgrid resolution is
increased, a larger portion of the spectrum is better resolved, cf. the
spectra obtained on $64^{3}$ and $96^{3}$ grid-points.
Moreover, the differences due to the use of the second order 
or the fourth order
accurate methods are expressed very clearly in the spectra; a strong 
reduction in energy
in the higher wavenumbers on the $32^{3}$ grid results when using the 
second order method.
This is consistent with the stronger attenuation of the high 
wavenumbers in the second
order method, compared to the fourth order accurate method.

\section{Concluding remarks}
\label{concl}

In the $\alpha$-framework we derived a new subgrid closure for the 
turbulent stress
tensor in LES, by using the Kelvin theorem applied to a filtered 
transport velocity. The
proposed \lesa subgrid model was shown to contain two other subgrid 
parameterizations,
i.e., the nonlinear gradient model and a model corresponding to the 
Leray regularization of
Navier-Stokes dynamics. Moreover, the \lesa model stress tensor 
contains an explicit
filtering in its definition, which sets it apart from other subgrid 
models in literature.
It was shown that this explicit filtering is essential for the \lesa 
models; without it,
the simulations develop a finite time instability. This instability 
was also observed in
an analysis of the viscous one-dimensional Burgers equation and 
appears to be associated
with the nonlinear gradient term.

The flow in a turbulent mixing layer was considered, in order to test 
the capabilities of
the three `nested' \lesa models. Through a comparison with dynamic 
(mixed) models, we
inferred that the Leray model provides particularly accurate 
predictions. The filtered
nonlinear gradient model in turn, compares well with the Leray model 
and corresponds
closely to the Bardina similarity model for the flows considered. The 
full \lesa model was
seen to generate too many small scales in the solution and 
correspondingly poorer
predictions, e.g., too large growth-rate, too high levels of 
turbulence intensities, etc.
An analysis of the resolved kinetic energy dynamics showed that the 
full \lesa model
contains two competing contributions which may tend to cancel and, 
thus, destabilize some
simulations involving this model. In particular, this tendency 
implied that simulations
using the full \lesa model were unstable in our simulations of 
turbulent mixing at high
subgrid resolution.

Apart from accuracy and a certain degree of numerical robustness, the 
computational
overhead associated with a subgrid model is an element of importance 
in evaluating
simulation methods. The computational effort associated with the 
three \lesa models is
considerably lower than that of dynamic (mixed) models. This is 
primarily a result of the
reduction in the number of explicit filtering operations required to 
evaluate the
\lesa model. Moreover, the accuracy of the predictions is higher for the
Leray model than for {\it any} of the other subgrid models 
considered. Consequently, the
Leray model is favored in this study and holds promise for 
applications to even more
complex and demanding flow problems. Preliminary results at 
significantly higher Reynolds
number suggest that the Leray model performs well also in this case.

Finally, we also considered the contribution to the dynamics arising 
from the spatial
discretization method at coarse subgrid resolution. In general, the 
role of numerical
methods in relation to LES has not yet been sufficiently clarified to determine
unambiguously whether the accuracy of predictions is restricted 
because of shortcomings in
the subgrid model, or whether this inaccuracy is due to spatial 
discretization effects.
Resolving this ambiguity and determining the main sources of error 
would help in finding
the best strategy for employing computational resources in an LES. In 
one strategy, a
grid-independent solution of the modeled LES equations at fixed 
filter-width is sought and
one only assigns computational resources for reducing numerical 
errors by increasing the
subgrid resolution ratio $\Delta/h$. This approach was applied here 
and used to evaluate
the additional dissipation that arises from the spatial finite volume 
discretizations at
either second order, or fourth order accuracy. This additional 
dissipation is associated
with the implicit filtering effect
of the small flow
features represented on the grid.


\end{document}